%% file: revised_unmarked_May2025.tex
\documentclass[sn-mathphys-num]{sn-jnl}

\input{macros}

\theoremstyle{thmstyleone}%
\theoremstyle{thmstyletwo}%

\theoremstyle{thmstylethree}%

\renewcommand{\P}{\mathbb{P}}

\begin{document}

\title[]{Inferring birth versus death dynamics for ecological interactions in stochastic heterogeneous populations}


\author[1]{\fnm{Erin} \sur{Beckman}}
\equalcont{}

\author[2]{\fnm{Heyrim} \sur{Cho}}
\equalcont{}

\author*[3]{\fnm{Linh} \sur{Huynh}}\email{linh.n.huynh@dartmouth.edu}
\equalcont{All authors contributed equally to this work and are listed in alphabetical order. }

\affil[1]{\orgdiv{Department of Mathematics and Statistics}, \orgname{Utah State University}, \orgaddress{\street{3900 Old Main Hill}, \city{Logan}, \postcode{84322}, \state{UT}, \country{USA}}}

\affil[2]{\orgdiv{School of Mathematical and Statistical Sciences}, \orgname{Arizona State University}, \orgaddress{\street{901 Palm Walk}, \city{Tempe}, \postcode{85281}, \state{AZ}, \country{USA}}}

\affil[3]{\orgdiv{Department of Mathematics}, \orgname{Dartmouth College}, \orgaddress{\street{29 N.~Main Street}, \city{Hanover}, \postcode{03755}, \state{NH}, \country{USA}}}

\abstract{
In this paper, we study the significance of ecological interactions and separation of birth and death dynamics in stochastic heterogeneous populations via general birth-death processes. Interactions can manifest through the birth dynamics, the death dynamics, or some combination of the two. The underlying microscopic mechanisms are important but often implicit in population-level data. We propose an inference method for disambiguating the types of interaction and the birth and death processes from population size time series data of a stochastic $n$-type heterogeneous population. The interspecies interactions considered can be competitive, antagonistic, or mutualistic. We show that different pairs of birth and death rates with the same net growth rate result in different time series statistics. Then, the inference method is validated in the example of a birth-death process inspired by the two-type Lotka-Volterra interaction dynamics. Utilizing stochastic fluctuations enables us to estimate additional parameters in this stochastic Lotka-Volterra model, which are not identifiable in a deterministic model.  }




\maketitle

\tableofcontents
\vspace{5mm}

\section{Introduction}\label{sec:introduction}
Complex systems are characterized by the relationship between macroscopic phenomena and the microscopic interactions between components.
In practice, collecting data on microscopic dynamics can be challenging, and it is easier to collect data on the overall population trends such as population sizes sampled at discrete times.
While population-level data provides an accessible and valuable overview of the system behaviors, understanding the microscopic processes that drive the observed trends is essential for a complete understanding of the system. 

This work focuses on understanding how individual-level interactions manifest in a process that involves birth and death dynamics, with motivation from ecological interactions. Extracting this information is important in many applications such as disambiguating exploitation versus interference competition in ecology, -static versus -cidal drug effects, evolution, and the hallmarks of cancers \cite{huynh2023inferring, huynh2022inference, jensen1987simple, pankey2004clinical, hanahan2011hallmarks, fouad2017revisiting}.

In this paper, we build on the work of Huynh et al.~2023 \cite{huynh2023inferring}, which focused on understanding birth versus death effects for homogeneous populations. Here, we extend the framework to heterogeneous populations with a particular emphasis on their ecological interactions. 
This extension is inspired by the work of Cho et al.~2023 \cite{cho2023designing}, which highlights the importance of different types of heterogeneous ecological interactions such as competitive, cooperative, and antagonistic. Interactions in heterogeneous populations have significant biological impact in various mechanisms including cancer therapy resistance \cite{meacham2013tumour,dagogo2018tumour,paczkowski2021}, antibacterial resistance \cite{dewachter2019bacterial,ogbunugafor2016competition,de2022microbial}, and ecological adaptations \cite{kendall2011demographic} such as bet-hedging \cite{levien2021non} and metapopulation migration \cite{doebeli1997evolution}.

The underlying model for our inference method is a general birth-death process. A birth-death process is a continuous-time Markov chain tracking the subpopulation sizes over time. 
The model is non-spatial and interactions between individuals are incorporated by allowing the birth and death rates to be dependent upon the population size in a nonlinear way.

In deterministic models, such as differential equation models of population growth, the way the system evolves is determined by the net growth rate, the birth rate minus the death rate. This means that there are infinitely many different pairs of birth and death rates with the same net growth rate and deterministic behavior. However, when using stochastic models, the properties of the system can vary as the birth and death rate change, even while the difference between the rates remains fixed. 

Other work has addressed the specific question of birth and death disambiguation, such as \cite{crawford2014estimation, roney2020estipop,gunnarsson2023statistical}. Crawford et al.~2014 \cite{crawford2014estimation} introduce an expectation-maximization algorithm to infer parameters in a general homogeneous birth-death process when the population size is sampled at discrete times.
Their method involves a likelihood, which requires explicitly the form of the birth and the death rates.
Here, we also assume that subpopulation sizes are sampled at discrete times, but unlike \cite{crawford2014estimation}, we require that the time between samples are sufficiently small. 
The method we present applies even when the functional form of the birth and death rates is initially unknown. 
Roney et al.~2020 \cite{roney2020estipop} use the maximum likelihood approach to estimate birth and death rates in a multi-type branching process and demonstrate their method in the case of linear birth and death rates. 
Their model does not include interaction between different populations. Gunnarsson et al.~2023 \cite{gunnarsson2023statistical} also study multi-type branching processes with a focus on inferring phenotype switching between the subpopulations. Unlike these prior studies, our method applies to general, nonlinear birth and death rates, which allows interactions between populations through these rates.

The remainder of the paper is structured as follows. In Section \ref{sec:ModelDefn}, we define the general birth and death process. We also define specifically the Lotka-Volterra birth-death process which will be used as an example throughout the paper. In Section \ref{sec:inference-data}, we describe the type of data which is required for the inference method and the simulation method which was used to generate the in-silico data. Section \ref{sec:significance} examines the significance of disambiguating the birth and death rates in the Lotka-Volterra example. The inference method is introduced in Section \ref{sec:inference}, applied to the example of the two-type Lotka-Volterra model, and the method is numerically validated in this case. We end the paper with a discussion, Section \ref{sec:conclusions}, including open questions which remain.

\section{Stochastic Models}\label{sec:ModelDefn}
\subsection{General Birth-Death Process}
We consider a population with $n$ distinguishable subtypes. Let $\boldsymbol{N}(t) = (N_1(t), N_2(t), \dots, N_n(t))$ be the number of individuals of each type at time $t$.
The population evolves as a continuous-time Markov process with the state space $\Z^n$, and state transitions happen according to exponential rates. The process is a birth-death process in which each transition results in the addition or removal of one individual. We will use the following notation throughout the paper:
\begin{itemize}
\item $\lambda_k(\bN(t))$ - the rate at which an individual of type $k$ is born into the system, 
\item $\mu_k(\bN(t))$ - the rate at which an individual of type $k$ is removed from the system, and 
\item $\eta_{total}(\boldsymbol{N}(t))$ - the rate at which the process transitions out of state $\boldsymbol{N}(t)$. It can be computed as the sum of all the transition rates out of state $\bN(t)$:

$$\eta_{total}(\boldsymbol{N}(t))= \sum_{k=1}^n \lambda_k(\boldsymbol{N}(t)) + \mu_k(\boldsymbol{N}(t)).$$
\end{itemize}
Let $\be_k$ be the $k$-th standard basis vector. For states $\boldsymbol{i}, \boldsymbol{j} \in \Z^n$, the process has the infinitesimal transition probabilities given by 
\begin{equation}
    \P(\boldsymbol{N}(t+\Delta t) = \boldsymbol{j}|\boldsymbol{N}(t) = \boldsymbol{i}) = \begin{cases} 1 -  \eta_{total}(\boldsymbol{N}(t))\Delta t + o(\Delta t) & \text{if }\boldsymbol{j} = \boldsymbol{i}\\
    \lambda_k(\boldsymbol{N}(t)) \Delta t + o(\Delta t) & \text{if }\boldsymbol{j} = \boldsymbol{i} + \be_k\\
    \mu_k(\boldsymbol{N}(t)) \Delta t + o(\Delta t) & \text{if }\boldsymbol{j} = \boldsymbol{i} - \be_k\\
        o(\Delta t) & \text{otherwise,}\\
    \end{cases}
\end{equation}
for $\Delta t$ sufficiently small. We require that $\lambda_k, \mu_k \geq 0$ and that $\mu_k = 0$ when $N_k = 0$. We also assume that the rates are such that the process does not explode in finite time (that is, that there are not an infinite number of transitions in a finite time interval). However, we do not require these rates to be linear, and the rate at which type $k$ gives birth or dies can depend on the number of individuals of any subtype. In this way, the subpopulations interact with each other through the birth and the death rates. Both intraspecies interactions and interspecies interactions are possible. Figure \ref{fig:alg-illustration} depicts a graphical representation when $n = 2$. See \cite{Allen} for a reference developing and discussing these types of processes in the context of biological applications or \cite{DurrettEoSP} for a more general introduction to continuous-time Markov chains. 

Throughout this paper, $o(\Delta t)$ is little-o notation and denotes the collection of any terms $f(\Delta t)$ which satisfy $f(\Delta t)/ \Delta t \rightarrow 0$ as $\Delta t \rightarrow 0$. 
\begin{figure}[H]
\begin{center}
\begin{tikzpicture}
\newlength{\myStep}
\setlength{\myStep}{16mm}
\begin{scope}
\clip (0.5\myStep,0.5\myStep) rectangle (3.5\myStep,3.5\myStep);
\draw[step = \myStep, very thick] (0mm,0mm) grid (4\myStep,4\myStep);
\end{scope}

\node(center) at (2\myStep, 2\myStep) {};
\node (top) at (2\myStep, 3\myStep) {};
\node (left) at (\myStep, 2\myStep) {};
\node (right) at (3\myStep, 2\myStep) {};
\node (bottom) at (2\myStep, \myStep) {};
\draw[fill] (center) circle (0.1\myStep);

\node at (0.5,2\myStep) { $i_2$};
\node at (2\myStep,0.5) { $i_1$};

\draw[-latex, thick] (0.5,3\myStep) -- +(0,0.5\myStep) node[left]  {$N_2$};
\draw[-latex, thick] (3\myStep,0.5) -- +(0.5\myStep,0) node[below] {$N_1$};

\draw[-latex,thick] (center) to[bend left = 60]  (top.south west) node[xshift=-7mm,yshift=-7mm,rotate=90] { $\lambda_2(i_1, i_2)$};
\draw[-latex,thick] (center) to[bend left = 60] node[midway,above] {$\lambda_1(i_1, i_2)$} (right);
\draw[-latex,thick] (center) to[bend left = 60] node[xshift=3mm,rotate=-90] {$\mu_2(i_1, i_2)$} (bottom);
\draw[-latex,thick] (center) to[bend left = 60] node[midway,below] {$\mu_1(i_1, i_2)$} (left);

\end{tikzpicture}
\end{center}

\caption{
    Graphical illustration of two-type birth-death processes. The arrows indicate possible state changes for the process. The exponential rates of each transition, listed by the arrows, are state-dependent. }
\label{fig:alg-illustration}
\end{figure}
There are many examples of common birth-death processes that fall within this general framework, such as linear growth, linear growth with immigration, logistic growth, and populations exhibiting the Allee effect.
\subsection{Two-Type Lotka-Volterra Birth-Death Process}\label{sec:model-LV}
Throughout the paper, we will focus on a particular example of a birth-death process that falls within this general framework where the birth and death rates are inspired by the Lotka-Volterra system of differential equations.

The Lotka–Volterra model is a classical model for two-species ecological interactions denoted here as $\bN(t) = (S(t), R(t)) $ and has the following form:
\begin{eqnarray}
\frac{dS}{dt} &=& r_SS\left( 1 -\frac{S}{K_S}-\alpha_S\frac{R}{K_S}  \right), \label{eqn:dSdt} \\ 
\frac{dR}{dt} &=& r_RR \left( 1 -\frac{R}{K_R} -\alpha_R\frac{S}{K_R} \right), \label{eqn:dRdt}
\end{eqnarray}
where the terms with $S^2$ and $R^2$ reflect intraspecies interactions and the terms with $RS$ and $SR$ reflect interspecies interactions. 
We choose notations $S$ and $R$ here as the two subpopulations could potentially represent drug-sensitive and drug-resistant cancer cells in applications \cite{paczkowski2021}. 
The parameters $r_S, r_R$ are the per capita intrinsic/low-density net growth rates of the $S$-individuals and $R$-individuals, $K_S, K_R$ represent the carrying capacities of $S$-individuals and $R$-individuals, and $\alpha_S$, $\alpha_R$ indicate how much the interspecies interactions affect the $S$-subpopulation and $R$-subpopulation, respectively.
The signs of $\alpha_S$, $\alpha_R$ indicate the type of interspecies interactions (e.g.~competition, cooperation, etc.). In particular, see Table \ref{tab:LVsign}. 

\begin{table}[h]
    \centering
    \begin{tabular}{|c|c|c|c|} \hline 
         & $\alpha_S > 0$ & $\qquad \alpha_S = 0 \qquad $& $\alpha_S < 0$  \\ \hline 
      $\qquad \alpha_R > 0\qquad $    & Competitive &  & S antagonizes R  \\ \hline 
      $\qquad \alpha_R = 0\qquad $    &  & Neutral & \\ \hline 
      $\qquad \alpha_R < 0\qquad $    & R antagonizes S &  & Mutualistic \\ \hline 
    \end{tabular}
    \caption{A summary of how the signs of the interaction parameters $\alpha_S$ and $\alpha_R$ in Equations (\ref{eqn:dSdt})--(\ref{eqn:dRdt}) determine the type of ecological interaction between the $S$-type and $R$-type subpopulations.}
    \label{tab:LVsign}
\end{table}
\noindent
The two-type deterministic Lotka-Volterra model typically takes the form in \eqref{eqn:dSdt} and \eqref{eqn:dRdt}, which captures the net growth of each population. It does not separate this growth into a birth rate and a death rate. To define an appropriate birth-death process that captures the intrinsic growth rate, the intraspecies interactions, and the interspecies interactions in the same way as the Lotka-Volterra system, each of these net effects must be split between birth behavior and death behavior. To do this, we introduce six parameters: $\delta_R,\delta_S, \gamma_R, \gamma_S, \sigma_R,$ and $ \sigma_S$. 

For the $S$-subpopulation equation, we split the net growth terms using $\delta_S, \gamma_S, \sigma_S$: 
\begin{align}
\text{intrinsic growth: \quad} &  r_S S = \underbrace{(1+\delta_S)r_S S}_{\text{birth rate}} - \underbrace{\delta_S r_S S}_\text{death rate}\\
\text{interspecies interaction: \quad} & -\frac{r_S}{K_S}S^2 = \underbrace{-\gamma_S \frac{r_S}{K_S} S^2}_{\text{birth rate}} - \underbrace{(1-\gamma_S)\frac{r_S}{K_S} S^2}_\text{death rate}\\
\text{intraspecies interaction: \quad} & -\alpha_S\frac{r_S}{K_S}SR = \underbrace{-\sigma_S \alpha_S \frac{r_S}{K_S} SR}_{\text{birth rate}} - \underbrace{(1-\sigma_S)\alpha_S\frac{r_S}{K_S} SR}_\text{death rate}.
\end{align}
The equations for $R$ are divided similarly using $\delta_R, \gamma_R, \sigma_R$. Increasing $\delta_S$ or $\delta_R$ increases the intrinsic birth and death rates while keeping the net intrinsic growth rate positive.
Similarly, the $\gamma_S, \gamma_R$ parameters divide the intraspecies interaction between birth and death rates, and the $\sigma_S, \sigma_R$ parameters split up the interspecies interaction. 

In the deterministic form, the equations of interest become
\begin{align}
\frac{dS}{dt} 
&= \!\begin{multlined}[t]\underbrace{\Big((1+\delta_S) r_S S-\gamma_S\frac{r_S}{K_S}S^2 -\sigma_S\alpha_S\frac{r_S}{K_S}RS\Big)}_{\text{birth rate}} \\ 
- \underbrace{\Big(\delta_Sr_SS + (1-\gamma_S)\frac{r_S}{K_S}S^2 + (1-\sigma_S)\alpha_S\frac{r_S}{K_S}RS \Big)}_{\text{death rate}},
\end{multlined}   \\ 
\frac{dR}{dt} &= \!\begin{multlined}[t] \underbrace{\Big( (1+\delta_R)r_RR - \gamma_R\frac{r_R}{K_R}R^2 -\sigma_R\alpha_R\frac{r_R}{K_R}SR \Big)}_{\text{birth rate}} \\
- \underbrace{\Big(\delta_Rr_R R + (1-\gamma_R)\frac{r_R}{K_R}R^2 + (1-\sigma_R)\alpha_R\frac{r_R}{K_R}SR\Big)}_{\text{death rate}}.
\end{multlined}
\end{align}
In the deterministic model, the new parameters are not identifiable because they cancel out; different values $\delta_R,\delta_S, \gamma_R, \gamma_S, \sigma_R,$ and $ \sigma_S$ do not result in different time series. Here we focus on a birth-death process to mirror these dynamics in a way that includes stochasticity. With $\delta_{S},\delta_{R} \in [0, \infty)$, $\gamma_S, \gamma_R, \sigma_S, \sigma_R \in [0,1]$, and $\alpha_S, \alpha_R \in \mathbb{R}$, each transition happens at a density-dependent rate given by 
\begin{align*}
    (S,R) \rightarrow (S+1,R)  &\hspace{.5in} \text{birth of $S$ at rate } \lambda_S(S,R)\\
    (S,R) \rightarrow (S-1,R)& \hspace{.5in}\text{death of $S$ at rate } \mu_S(S,R)\\
    (S,R)\rightarrow (S,R+1)& \hspace{.5in}\text{birth of $R$ at rate } \lambda_R(S,R) \\ 
    (S,R) \rightarrow (S,R-1)& \hspace{.5in} \text{death of $R$ at rate } \mu_R(S,R), 
\end{align*}
where each of these rates is defined to be
\begin{align}
    \lambda_S(S,R) &= 
        \max\Big\{(1+\delta_S) r_S S-\gamma_S\frac{r_S}{K_S}S^2  - \sigma_S\alpha_S\frac{r_S}{K_S}RS, 0 \Big\}\label{eq:BSrate}\\
    \mu_S(S,R) &= \max\Big\{\delta_Sr_SS + (1-\gamma_S)\frac{r_S}{K_S}S^2 + (1-\sigma_S)\alpha_S\frac{r_S}{K_S}RS,0 \Big\}\label{eq:DSrate}\\
    \lambda_R(S,R) &= \max\Big\{(1+\delta_R)r_RR - \gamma_R\frac{r_R}{K_R}R^2 -\sigma_R\alpha_R\frac{r_R}{K_R}SR  ,0\Big\}\label{eq:BRrate}\\
    \mu_R(S,R) &= \max\Big\{\delta_Rr_R R + (1-\gamma_R)\frac{r_R}{K_R}R^2 + (1-\sigma_R)\alpha_R\frac{r_R}{K_R}SR,0\Big\}.\label{eq:DRrate}
\end{align}
The max in each of these rates ensures that the rates are always nonnegative.

\section{Data Description} \label{sec:inference-data}
Our dataset consists of time series of the subpopulation counts at different times. We assume that the data is sampled at discrete time points and therefore we do not have access to the number of births and the number of deaths occurring in a time interval. Only the net population change over a time interval is known.

\paragraph{Stochastic Simulation} 
To study the effect of interactions happening in the birth and death rates and to demonstrate the inference method on the Lotka-Volterra birth-death process described above, we generate several time series in-silico for analysis. 

We use the tau-leaping algorithm to simulate the population time series. In this algorithm, the birth-death process is approximated by updating the birth and death rates at deterministic intervals of size $\tau > 0$, rather than after each birth and death event. With this approximation, all events of a particular type occurring in the $\tau$-interval are simulated together using a Poisson random variable, rather than each event being tracked individually. This is an established stochastic simulation algorithm which speeds up computation of the time series while still maintaining reasonable accuracy. The error associated with the tau-leaping method is of order $\tau$; see \cite{warne2019simulation} for an analysis of this method and comparison to other simulation algorithms. In generating the data for analysis, we used $\tau =0.1$. Using the tau-leaping scheme has the added benefit of creating data which is of the same form as datasets typically collected in laboratories \cite{emond2023cell, maltas2024drug, maltas2024frequency, roshan2015dynamic} or clinics. In these settings, it can be difficult to track individual cells in experiments or patients and instead, population counts are updated at fixed time intervals.
\paragraph{Parameter Choices} We focus on parameters relevant to two different types of prostate cancer cell lines, PC3 and DU145. Parameter values for these cancer lines have been identified in previous studies \cite{paczkowski2021} and are listed in Appendix \ref{secA1}. We also include simulations with artificial parameter choices to study different interaction regimes. Each numerical study will indicate the parameter choices that were made, either explicitly or by reference to the cell line name.

\paragraph{Monoculture and coculture experiments}
To infer all model parameters in the Lotka-Volterra birth-death process in Section \ref{sec:inferencealgorithm_LV}, we use the sequential inference approach as in \cite{cho2023designing}, which involves two types of experiments: monoculture and coculture.  The difference between these two types of data is the initial conditions. To produce monoculture data, we begin with an initial condition with only one subtype present, either $S$ or $R$ type.
Notice that there is no mutation in this model, so if the initial condition only contains one subpopulation, then there is at most one subpopulation present for all time. 
To produce coculture data, the experiment begins with both types present. We require that each subpopulation size is able to be counted separately throughout the experiment. In-vitro, this can often be done using a fluorescence technique \cite{paczkowski2021}.

\section{Biological Significance of Birth versus Death Interaction Regulation}\label{sec:significance}

In this section, we examine the effects of birth versus death regulation of intraspecies or interspecies interactions in the Lotka-Volterra process described in Section \ref{sec:model-LV}. We study the significance of the intraspecies competition parameter, $\gamma_k \in [0,1]$, and the interspecies interaction parameter, $\sigma_k \in [0,1]$, by conducting a parameter sensitivity analysis of $\gamma_k$ and $\sigma_k$ on various population properties, including variance in the subpopulation sizes and the probability that the resistant subpopulation survives for a fixed period of time. We demonstrate that there are quantitative differences in the population dynamics when the intraspecies and interspecies regulations are changed from affecting the birth process to the death process. 

\subsection{Time Series Statistics}\label{sec:trajstats}
\label{sec:trajstats}
\begin{figure}[h]
    \centerline{ \footnotesize{ (a) PC3: $\gamma_R=0,\sigma_R=0$,\hspace{1.2cm} $\gamma_R=0,\sigma_R=1$,\hspace{1.2cm} $\gamma_R=1,\sigma_R=0$,\hspace{1.2cm} $\gamma_R=1,\sigma_R=1$ \hspace{0.5cm} } }
    \centerline{ 
\includegraphics[width=16cm]{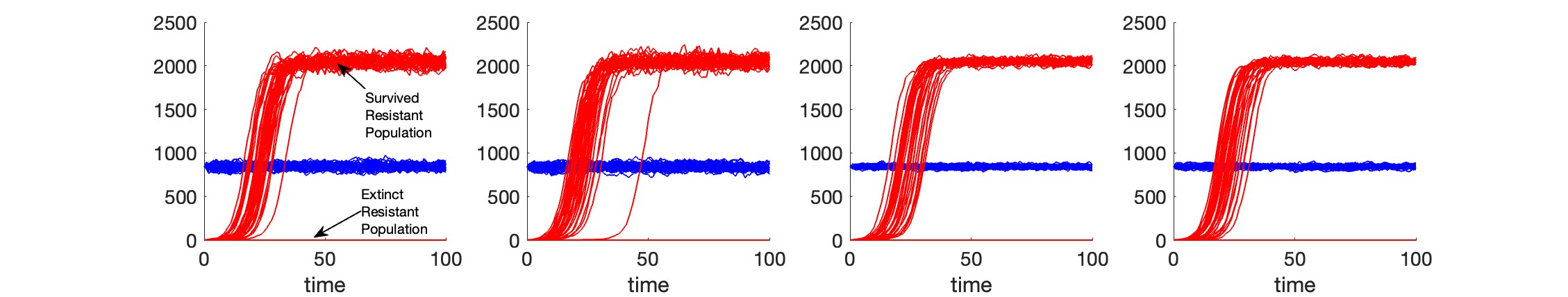}
}
    \centerline{ \footnotesize{ (b) DU145: $\gamma_R=0,\sigma_R=0$,\hspace{1.2cm} $\gamma_R=0,\sigma_R=1$,\hspace{1.2cm} $\gamma_R=1,\sigma_R=0$,\hspace{1.2cm} $\gamma_R=1,\sigma_R=1$ \hspace{0.8cm} } }
    \centerline{ 
\includegraphics[width=16cm]{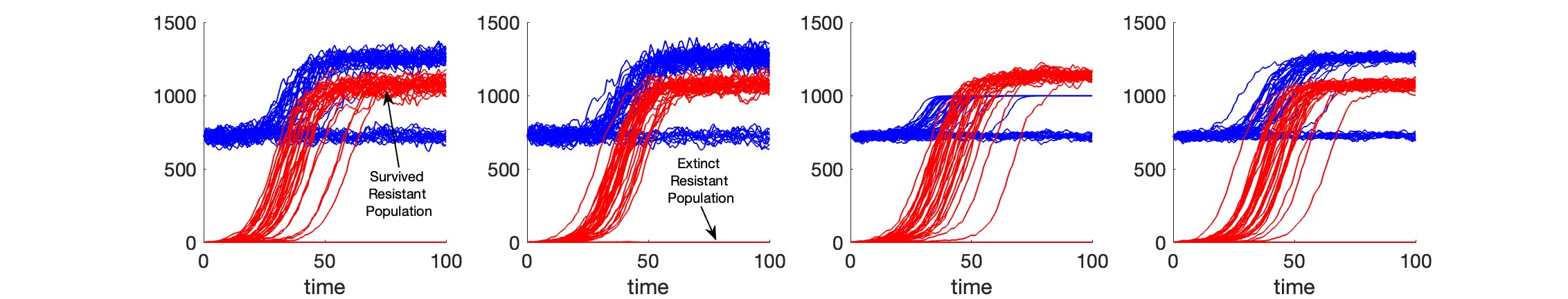}
}
\caption{ 
In-silico time series of sensitive (blue) and resistant (red) populations of (a) PC3 and (b) DU145 cell lines for different $\gamma_R$ and $\sigma_R$ values. The variance is most affected by $\gamma_R$, i.e.,  the regulation of intraspecies competition on either birth or death. 
}

    \label{fig:traj}
\end{figure}

First, we examine the overall time series of the subpopulations in the PC3 and DU145 cell lines and demonstrate that the regulation of intraspecies competition via $\gamma_k$ affects the variation within the time series. Figure \ref{fig:traj} shows the time series obtained from Lotka-Volterra birth-death process simulations, where the blue lines represent the sensitive population and the red lines represent the resistant population. It shows the extreme cases of $\gamma_k\in\{0,1\}$ and $\sigma_k \in\{ 0,1\}$. For simplicity, we set $\gamma_S=\gamma_R$ and $\sigma_S=\sigma_R$ in all the simulations displayed in Figure \ref{fig:traj}. The choice of $\gamma_S=\gamma_R=1$ is when the intraspecies competition regulates the birth process only, and when $\gamma_S=\gamma_R=0$, the intraspecies competition is entirely in the death process. When the interspecies interaction regulates the birth only, we have $\sigma_S=\sigma_R=1$, and the case $\sigma_S=\sigma_R=0$ is when the interspecies interaction is only present in the death process. 
The simulations are initiated with $[S_0,R_0] = [K_S-1, 1]$, assuming that one cell from the sensitive population has mutated to a resistant cell. For these parameter values, this gives an initial condition of $[S_0,R_0] = [842,1]$ and $[S_0,R_0] = [723,1]$ for PC3 and DU145, respectively. The time series in Figure \ref{fig:traj} of each population can be separated into two groups--the resistant cells either growing to full capacity or the resistant cells going extinct. 

In the case of PC3 in Figure \ref{fig:traj}(a), the capacity of sensitive cells is the same regardless of whether or not the resistant cells survive because the interaction parameter $\alpha_S$ is close to being trivial, $\alpha_S \approx 0$. With the DU145 parameters, we see in Figure \ref{fig:traj}(b) that the sensitive cells of DU145 either stay at the sensitive cell carrying capacity, corresponding to the runs when the resistant cells go extinct, or they reach a higher level when the resistant cells survive. This increased carrying capacity is due to the interaction type where the resistant population boosts the sensitive cells, $\alpha_S < 0$. 
The third panel for the DU145 case shows a phenomenon that if the resistant population survives, then the number of sensitive cells eventually becomes constant. This is because at this point, the birth rate and death rate of the sensitive cells are both zero. In particular, though the choice of $\sigma_k$ and $\gamma_k$ does not affect the net growth rate of the subpopulations for most parameter values, if the maximum in Equations \eqref{eq:BSrate}-\eqref{eq:DRrate} forces a birth or death rate to be zero, then the net growth rate can be impacted by the choice of $\gamma_k$ or $\sigma_k$. In this case, the net growth rate becomes zero in a range where other choices of $\gamma_k$ and $\sigma_k$ would have a nonzero net growth rate.  

\begin{figure}[!tb]
    \centerline{ \footnotesize{ (a) PC3 : $\alpha_S=0.027, \alpha_R=0.159$ \hspace{7.5cm} } }
    \centerline{ 
\includegraphics[width=15cm]{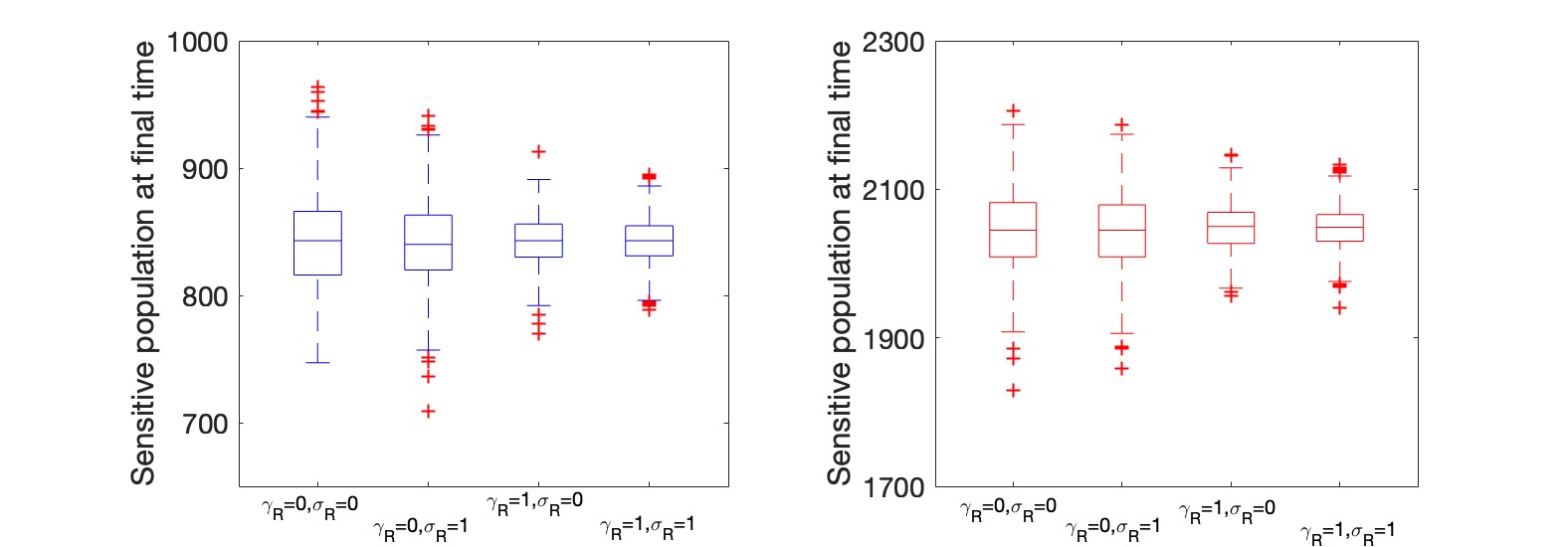}
}
    \centerline{ \footnotesize{ (b) DU145 : $\alpha_S=-0.501, \alpha_R=0.221$ \hspace{6.8cm} } }
    \centerline{ 
\includegraphics[width=15cm]{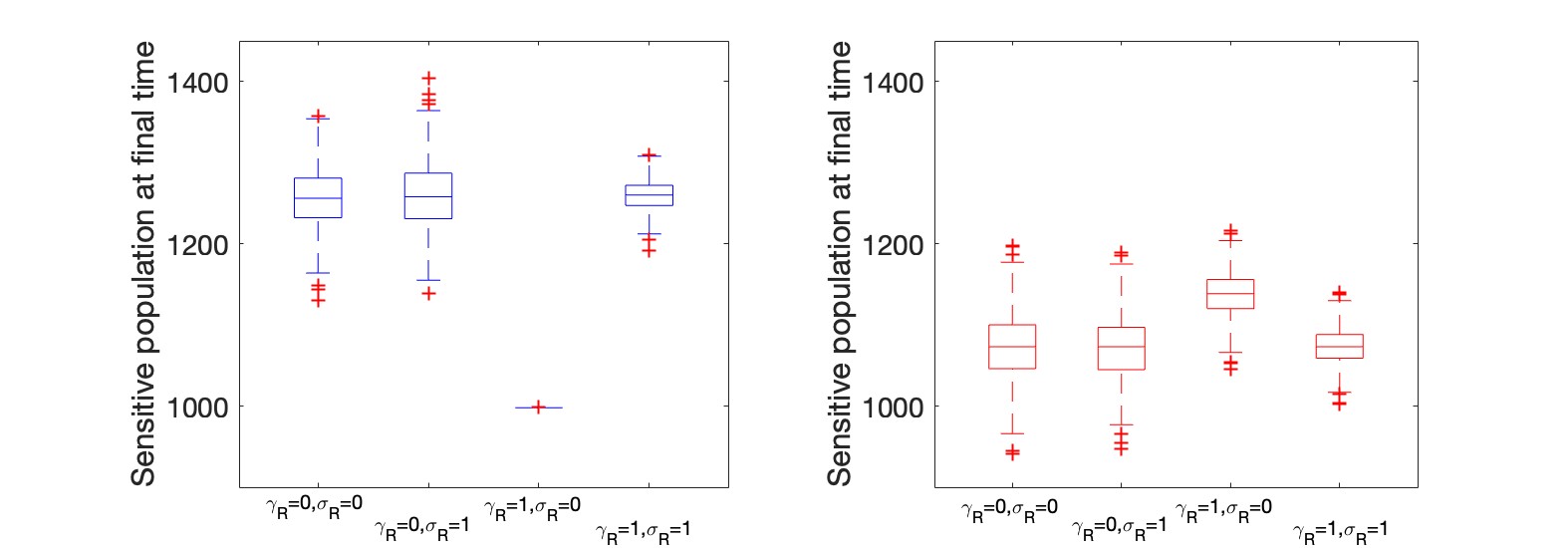}
}
\caption{ 
Box plot statistics of sensitive (left) and resistant (right) populations of (a) PC3 and (b) DU145 cell lines at the final simulation time $t=100$, given that the resistant cells survive. The range of the number of cells at the final time is larger when $\gamma_R=0$ compared to $\gamma_R=1$, that is, when the intraspecies competition regulates death rather than birth. In the case of DU145, the median of the population differs from the other cases when $\gamma_R=1$ and $\sigma_R = 0$. 
}

    \label{fig:end_stat} 
\end{figure}

Figure \ref{fig:end_stat} shows a box plot of the subpopulation sizes at time $t=100$, given that the resistant cells survived until time $t = 100$. The third box plot ($\gamma_R = \gamma_S = 1$, $\sigma_R = \sigma_S = 0$) for DU145 shows again the effect of the sensitive cells reaching a point where the birth rate and the death rate are zero. This affects both the median and the range of the sensitive cells. Because the sensitive cell population is cut off at a different point in this parameter range, it also affects the median of the resistant cell population. 

In the other cases, the medians of the time series in each group do not change substantially depending on the choice of $\gamma_k$ and $\sigma_k$. However, the size of the interquartile range of the time series changes as $\gamma_k$ is varied. In particular, the interquartile range of the final subpopulation sizes decreases when $\gamma_R$ changes from $0$ to $1$. This is observed in both PC3 and DU145 cell lines at both extremes of $\sigma_k$. This effect is more extreme in the PC3 cell line. From this, we can see that as the intraspecies competition moves to decreasing the birth rate, rather than increasing the death rate, the range of the final population size decreases. Changing $\sigma_R$, the interspecies interaction, from the death to the birth rate does not have as large of an impact on the range of the final population sizes. 

\subsection{Computational Survival Probability} \label{sec:survivalprob}
In this section, we study the effect of birth- versus death-regulated interactions on the survival of the resistant population for different interspecies interaction regimes. We compute the probability that the resistant population does not go extinct for a fixed time period of $T = 100$ when the population begins with one resistant cell. We refer to this as the survival probability. This quantity demonstrates how likely the resistant population is to grow to carrying capacity, given that it begins with a single cell.
The survival probability is numerically computed from ${M=}$ 10000 time series by computing the proportion of runs in which the resistant population survived until time $T=100$. The simulation is initiated with $[S_0,R_0] = [K_S-1, 1]$ as in Section \ref{sec:trajstats}. 
The proportion of the $M$ runs with survival at time $T$ is the approximate survival probability:  
\begin{align}
P\left[ R(T) \neq 0 | S(0)=K_S-1, R(0) = 1\right] \approx \frac{1}{M} \sum_{i=1}^M \mathbbm{1}_{ R^i(T) \neq 0},
    \label{eq:survprob}
\end{align}
Here, $R^i$ is the resistant population in the $i$-th simulation and $T$ is our final simulation time, taken as $T=100$.

\begin{figure}[!b]
    \centerline{ \footnotesize{ (a) PC3 ($\alpha_S=0.027, \alpha_R=0.159$) \hspace{1.7cm} (b) DU145 ($\alpha_S=-0.501, \alpha_R=0.221$) } } 
    \centerline{ 
\includegraphics[width=6cm]{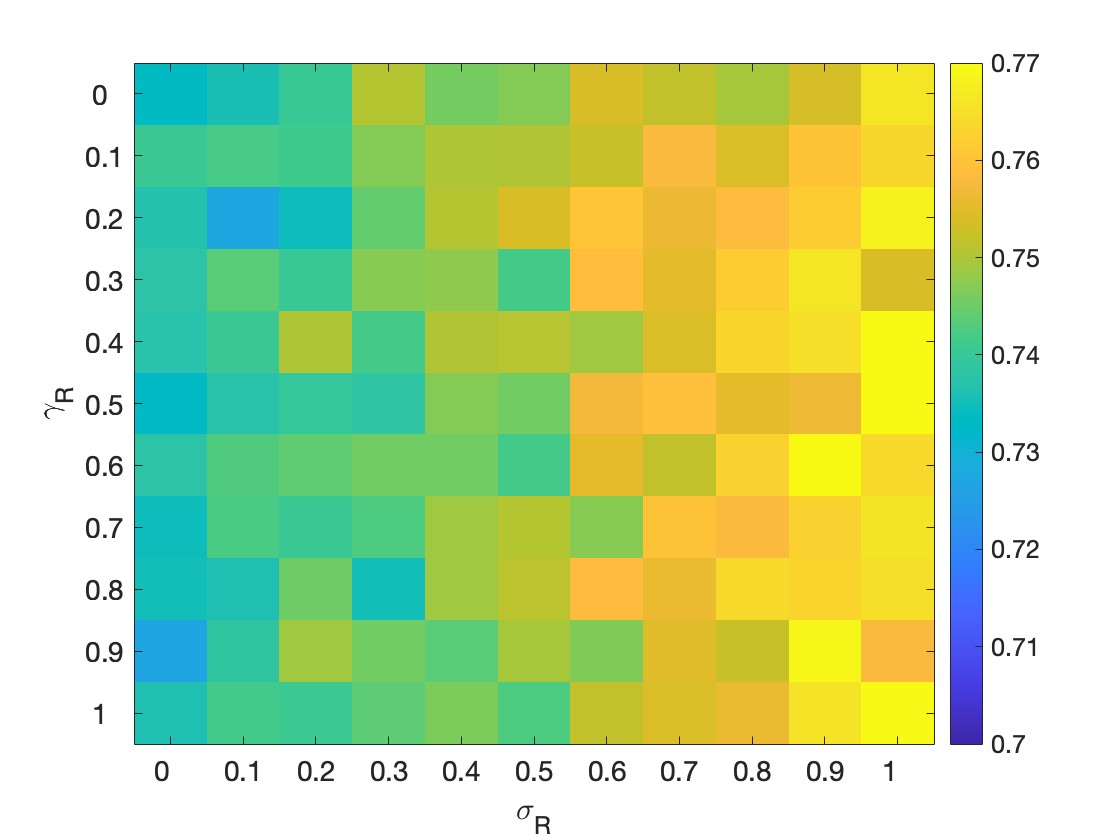}
\includegraphics[width=6cm]{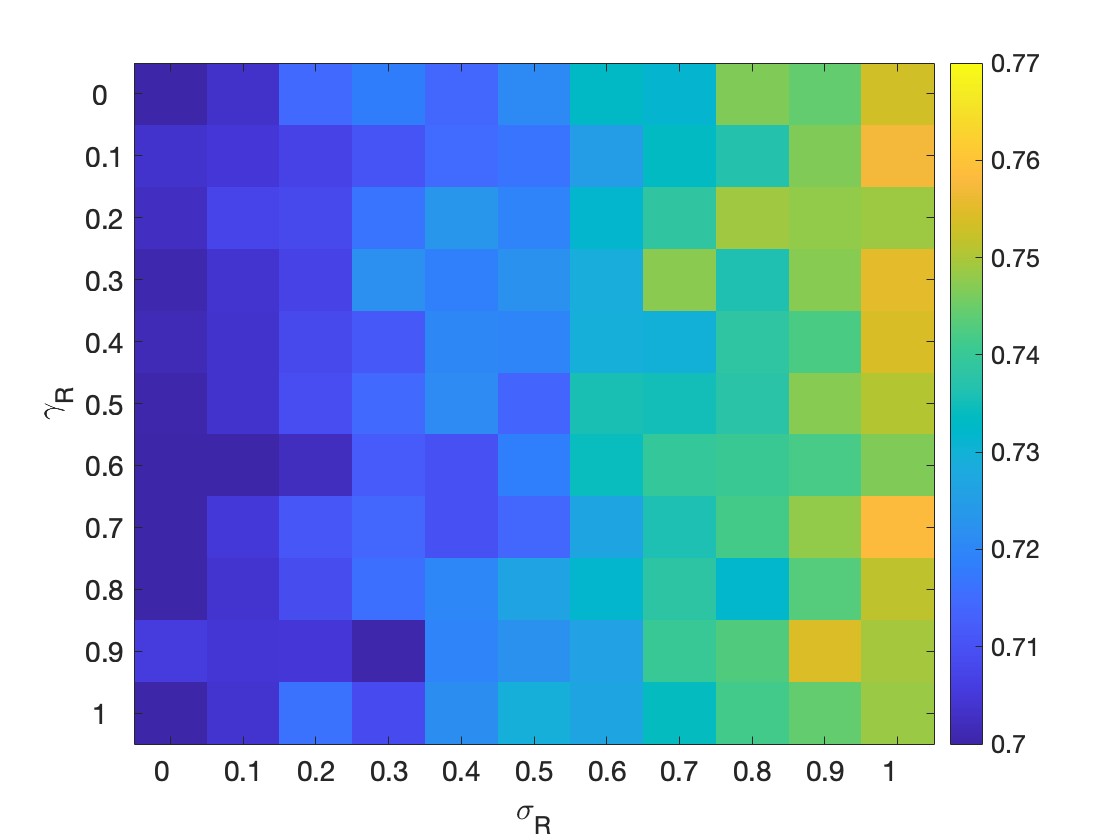}
}
    \centerline{ \footnotesize{ (c) case 3 ($\alpha_S=0.159, \alpha_R=0.027$) \hspace{1.9cm} (d) case 4 ($\alpha_S=0.221, \alpha_R=-0.501$)} }
    \centerline{ 
\includegraphics[width=6cm]{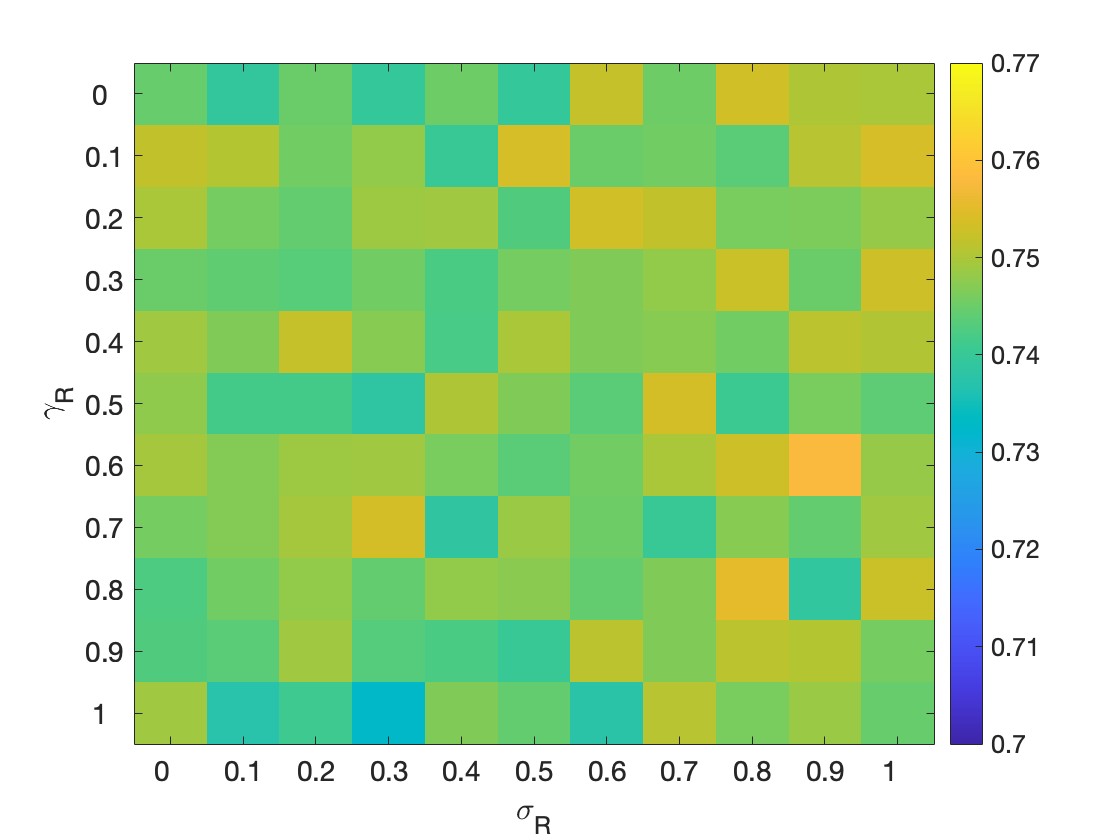}
\includegraphics[width=6cm]
{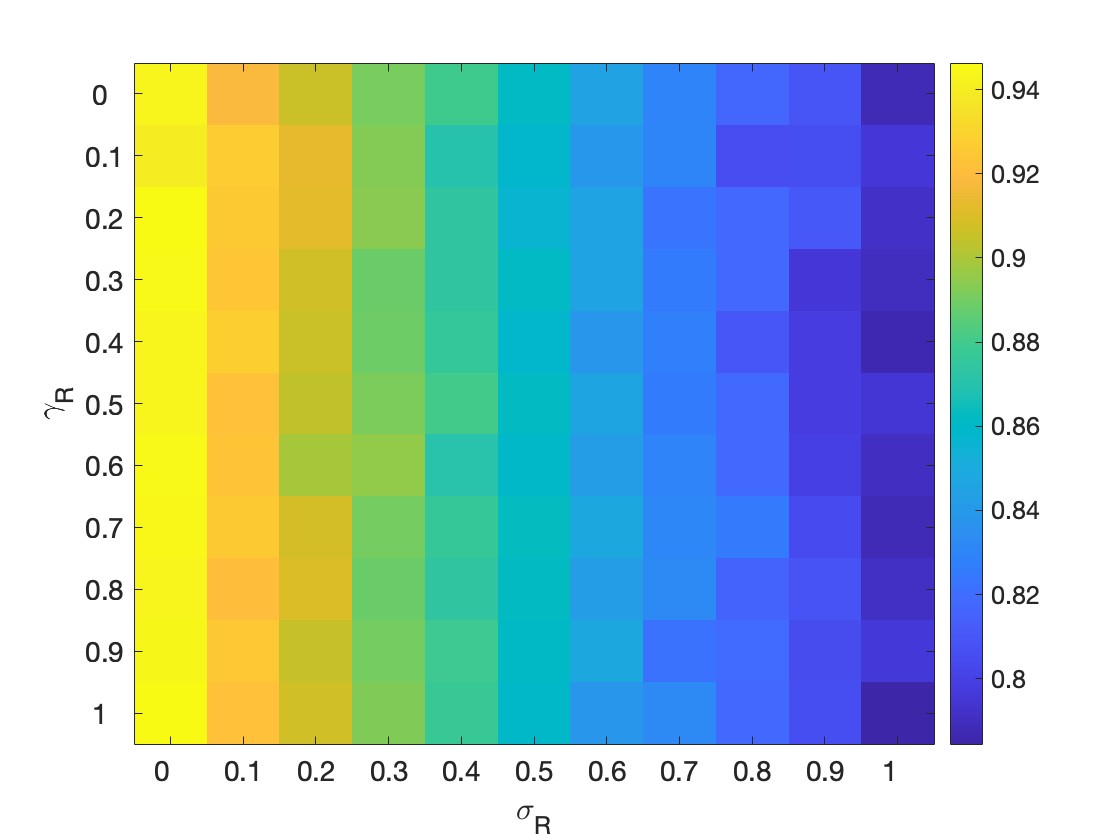}
}
\caption{
Survival probability in Equation \eqref{eq:survprob} with respect to $\gamma_R$ and $\sigma_R$ in various interaction cases (a-d). 
The survival probability is sensitive to $\sigma_R$, but not to $\gamma_R$. It implies that the birth versus death regulation of interspecies interactions impacts the survival probability more than the intraspecies competition term. 
 }

    \label{fig:SurvProb_PC3}
\end{figure}

\begin{figure}[!tb]
    \centerline{ 
\includegraphics[width=14cm]{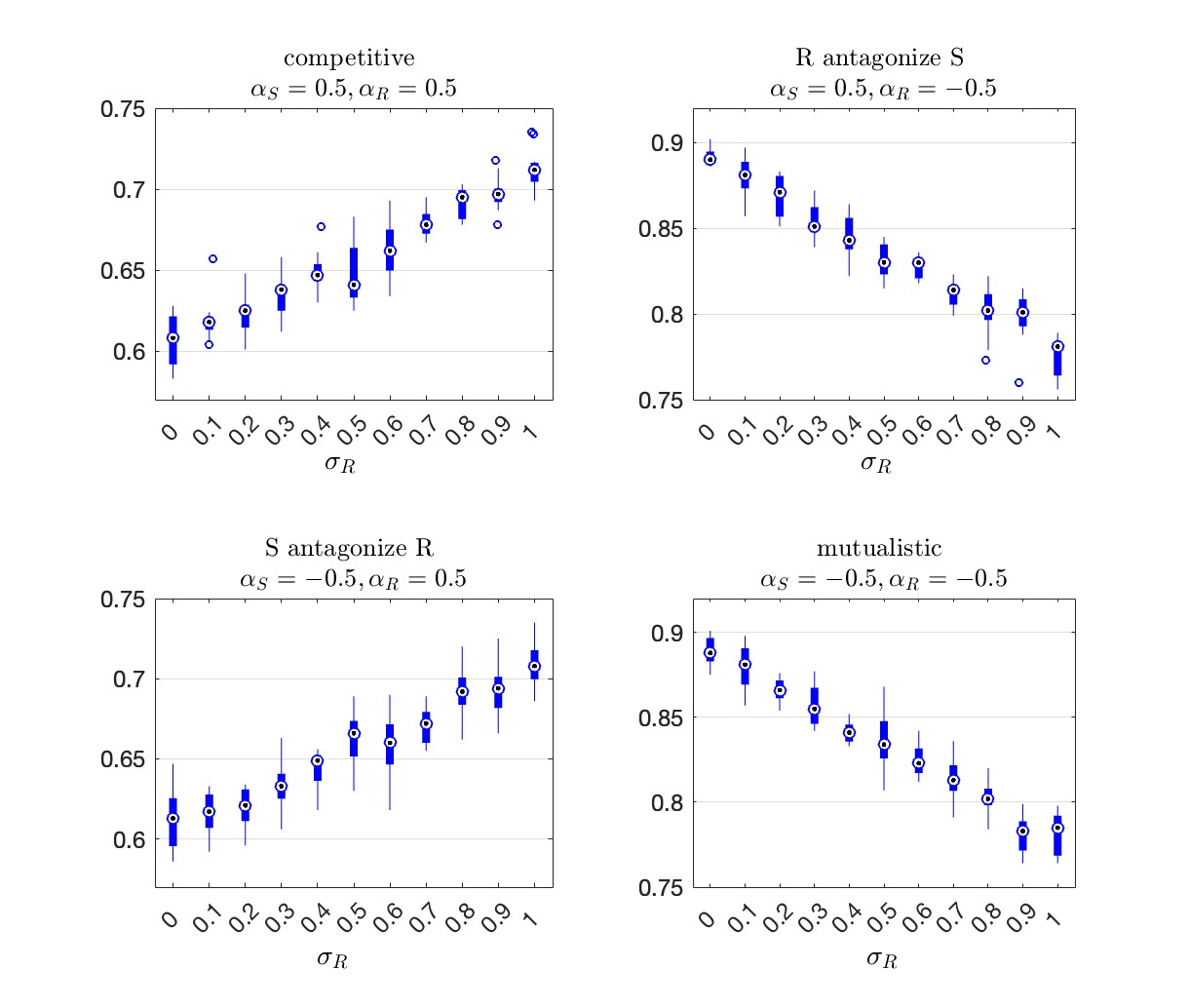}
}
\caption{
Survival probability of the resistant population with respect to $\sigma_R$ for different interaction regimes, including competitive, mutualistic, and antagonistic interactions ($\alpha_{S} = \pm 0.5, = \alpha_{R} = \pm 0.5$). When $\alpha_{R} = 0.5$, the sensitive cells negatively impact the resistant cells, so the survival probability increases as $\sigma_R$ increases. When $\alpha_{R} = -0.5$, the sensitive cells positively impact the resistant cells, so the survival probability decreases as $\sigma_R$ increases. 
}

    \label{fig:SurvProb_PCv2}
\end{figure}

Figure \ref{fig:SurvProb_PC3} plots the survival probability of the resistant population for different values of $\gamma_R = 0, 0.1,...,1$ and $\sigma_R = 0, 0.1,...,1$. In the cases of PC3 and DU145, the survival probability is more sensitive to $\sigma_R$ compared to $\gamma_R$. The birth versus death regulation of interspecies interaction affects the survival probability more than the intraspecies competition. The survival probability in both PC3 and DU145 increases about 7\% as $\sigma_R$ increases from 0 to 1. Both cases are when $\alpha_R>0$. In addition, we test two different types of interaction where $\alpha_R$ has different signs, $\alpha_R \approx 0$ in Figure \ref{fig:SurvProb_PC3}(c) and  $\alpha_R < 0$ in Figure \ref{fig:SurvProb_PC3}(d). When $\alpha_R \approx 0$ the effect of the sensitive cells on the resistant cells is small, thus $\sigma_R$ is not applicable and the survival probability stays relatively consistent regardless of $\sigma_R$. However, when $\alpha_R<0$ in Figure \ref{fig:SurvProb_PC3}(d), the survival probability decreases as $\sigma_R$ increases. Thus, the sign of $\alpha_R$ changes the sign of the correlation between the survival probability and $\sigma_R$.

We further study the survival probability of the four interspecies interaction types--competitive ($\alpha_S=0.5, \alpha_R=0.5$), antagonistic ($\alpha_S=-0.5, \alpha_R=0.5$ and $\alpha_S=0.5, \alpha_R=-0.5$), and mutualistic ($\alpha_S=-0.5, \alpha_R=-0.5$). The parameters other than $\alpha_i$ are chosen to be the same as PC3. 

We again confirm that the correlation of survival probability and $\sigma_R$ switches depending on the sign of $\alpha_R$. When $\alpha_R=0.5>0$ (competitive interactions and when S antagonizes R), the survival probability increases as $\sigma_R$ increases (that is, the survival probability is higher when the interaction regulates the birth process rather than the death process). 
However, when $\alpha_R=-0.5<0$ (when R antagonizes S and mutualistic interactions), the survival probability decreases as $\sigma_R$ increases, so the survival probability is higher when the interaction regulates the death process rather than the birth process. The value of $\sigma_S$ does not affect the survival probability of the resistant population. 

Since the regulation of interspecies interaction affects the survival probability more than the regulation of intraspecies competition does, for fixed $\alpha_R$, we compute the summary statistics across $\sigma_R$ for a fixed $\gamma_R$. Figure \ref{fig:SurvProb_PCv2} shows the box plot of the survival probabilities with respect to $\sigma_R$ for different interaction types.

The change in the survival probability of the resistant population across different values of $\sigma_R$ depends on the sign of $\alpha_R$ but not on the sign of $\alpha_S$. It is reasonable that the parameter that describes how the sensitive cells impact the resistant cells affects the survival probability of the resistant population more than the type of interaction in the opposite direction. When $\alpha_R=0.5$ and the resistant cells compete with the sensitive cells, the survival probability is around $0.55$ to $0.7$. However, when $\alpha_R=-0.5$ and the sensitive cells positively impact the resistant cells, the survival probability increases to the level of $0.75$ to $0.95$. In addition to the numerical probability of survival, we again confirm that the correlation of $\sigma_R$ to the survival probability changes depending on the sign of $\alpha_R$. When $\alpha_R>0$, the survival probability increases as $\sigma_R$ increases, but when $\alpha_R<0$, the survival probability decreases as $\sigma_R$ increases.

\section{Inference Methods}\label{sec:inference}
Section \ref{sec:significance} demonstrates the significance that varying the birth and death rate pairs can have on the population properties. Now we describe an inference method to disambiguate these rates given discrete time series data, as outlined in Section \ref{sec:inference-data}. This method is a generalization of the work of Huynh et al. 2023 \cite{huynh2023inferring} to the heterogeneous population setting. 
\subsection{Mathematical Theory}\label{sec:inference-theory}
We begin the discussion of the inference method by outlining the mathematical basis for the technique. In a birth-death process, one can approximate the number of births and the number of deaths for each species in a short time interval as independent Poisson random variables, as described in Section \ref{sec:inference-data}. Such an approximation uses the birth and death rates for each species at time $t$ to update the population increase and decrease over entire interval $[t, t+\Delta t)$. The approximation ignores any rate updates that happen in this interval. This is the approximation made when implementing a tau-leaping simulation of a birth-death process. The error in the expected population size at a fixed time decreases linearly with $\Delta t$ \cite{warne2019simulation}. Therefore, for this approximation and the resulting inference technique to be valid, we will assume throughout that $\Delta t$ is small. 

Again, using the notation that $N_k(t)$ is the number of type $k$ individuals alive at time $t$, we define the change in each type over a $\Delta t$ time period by
\begin{align}
\Delta N_k(t) &:= N_k(t + \Delta t) - N_k(t) \quad \text{for } k = 1, 2, \dots, n.
\end{align}
Letting $\Delta^+ N_k(t)$ be the number of births in the time interval $(t, t+\Delta t)$ and $\Delta^- N_k(t)$ be the number of deaths in that interval, 
\begin{align}
\Delta N_k(t) = \Delta^+ N_k(t) - \Delta^- N_k(t).
\end{align}
Conditioned on knowing $\bN(t)$, the distribution of each of these is approximately Poisson: 
\begin{align}
    \Delta^+ N_k(t)|\bN(t) &\stackrel{d}{\approx} \Poi\left(\lambda_k(\bN(t)) \Delta t\right)\label{eq:BBin}\\ 
    \Delta^- N_k(t)|\bN(t) &\stackrel{d}{\approx} \Poi\left(\mu_k(\bN(t)) \Delta t\right). \label{eq:DBin}
\end{align}
From this, the expected value and variance of $\Delta N_k(t)|\boldsymbol{N}(t)$ can be approximated, up to $o(\Delta t)$, as 
\begin{align}
\E[\Delta N_k(t)|\bN(t)] \approx \Delta t (\lambda_k(\bN(t)) - \mu_k(\bN(t))), \label{eq:mean-DeltaN}\\
\Var(\Delta N_k(t)|\bN(t)) \approx \Delta t(\lambda_k(\bN(t))+\mu_k(\bN(t))). \label{eq:var-DeltaN}
\end{align}

For each species $k$, this results in two equations \eqref{eq:mean-DeltaN} and \eqref{eq:var-DeltaN} and solving for $\lambda_k, \mu_k$ gives the equations
\begin{align}
    \lambda_k(\bN(t)) &\approx \frac{ \Var(\Delta N_k(t)|\bN(t))+\E[\Delta N_k(t)|\bN(t)]}{2\Delta t}\label{eq:birthratefromEV}\\
    \mu_k(\bN(t)) &\approx \frac{\Var(\Delta N_k(t)|\bN(t)) - \E[\Delta N_k(t)|\bN(t)]}{2 \Delta t}.
    \label{eq:deathratefromEV}
\end{align}
The rates depend on the entire state of the system $\bN(t)$, as the variance and expected value in \eqref{eq:birthratefromEV}, \eqref{eq:deathratefromEV} are conditional on $\bN(t)$, but they do not require the functional form of $\lambda_k, \mu_k$.

\subsection{Birth and Death Rate Inference Algorithm  }\label{sec:inferencealgorithm_general}
    We propose an inference method to approximate the birth and the death rate in a general, interacting birth-death process using the mean and variance conditioned on different population points $\bN = (N_1, \dots, N_n)$. For clarity of notation, we write the algorithm process for $n = 2$. However, this method can be generalized to $n > 2$ to allow for more distinguishable subtypes. 
    
Fix a $\Delta t > 0$ and a dataset of time series of the form $\bN = (N_1(\bt), N_2(\bt))$ for some time vector $\bt = \{t_s\}$. This $\bt$ can be different for each time series in the dataset. We use the notation $(N_1(t_s), N_2(t_s))$ to refer to a generic point in a time series and call it a population point. 
   
\begin{enumerate}
    \item  First, we divide the space $\R^{\geq 0} \times \R^{\geq 0}$ into a grid. The grid can have a fixed grid size $\Delta x$ or can be chosen to vary, to more appropriately capture the data.    
    \item We then map each time series onto this grid, ignoring the time variable and placing a population point at $(N_1(t_s), N_2(t_s)) \in \R^2$ for all $t_s$ in the time vector $\bt$.
    \item For each grid block with midpoint $\bN^{i,j} = (N_1^{i,j}, N_2^{i,j})$, we approximate $\E[\Delta \bN|\bN^{i,j}], \Var(\Delta \bN|\bN^{i,j})$ with the following steps. 
    \begin{enumerate}
        \item Each population point $(N_1(t_s), N_2(t_s))$ in the $(i,j)$-th grid block has an associated $(N_1(t_s + \Delta t), N_2(t_s+\Delta t))$ in its time series. If required by the time spacing in the time series, this value can be interpolated. Note that the time series and the time $t_s$ may be different for each population point in the grid block. From these values we can compute $\Delta \bN^p$ for each population point $p$ in the block. Here we suppress the dependence upon $t$ since all these values are grouped together based solely on their spatial position, disregarding their temporal position. 
        \item With this collection of $\{\Delta \bN^p\}$ for each population point $p$ in the grid block, we compute the sample mean and variance of $\Delta \bN$ for the grid block. 
        \item We associate this mean and variance to the grid block midpoint to approximate  $\E[\Delta \bN | (N_1^{i,j}, N_2^{i,j}) ]$ and $\Var(\Delta \bN | (N_1^{i,j}, N_2^{i,j}) )$. 
    \end{enumerate}
    \item After completing this process for every grid block, we have an approximate value of $\E[\Delta \bN | (N_1^{i,j}, N_2^{i,j}) ]$ and $\Var(\Delta \bN | (N_1^{i,j}, N_2^{i,j}) )$ for each grid block midpoint. We then use equations \eqref{eq:birthratefromEV} and \eqref{eq:deathratefromEV} to approximate $\lambda_k(N_1^{i,j}, N_2^{i,j}), \mu_k(N_1^{i,j}, N_2^{i,j})$ at each grid block midpoint. 
\end{enumerate}
The choice of grid size is important. For the method to be accurate, there needs to be enough data points in each grid block to get a good approximation of the mean and variance, so the grid blocks must be big enough to have a large number of points in each nonempty grid block. However, because $\lambda_k, \mu_k$ are functions of $\bN$, choosing the grid too large may group points together which have quite different rates. This will also reduce the accuracy of the method. 
In the homogeneous population case, the analysis of Huynh et al.~2023 \cite{huynh2023inferring} suggests that intermediate grid sizes are optimal. For the two-population Lotka-Volterra model, we analyze the error as a function of grid size $\Delta x$ in Figure  \ref{fig:error_dS}. 

\subsection{Inference Method for Lotka-Volterra Birth-Death Process}\label{sec:inferencealgorithm_LV} 

We demonstrate the general inference method described above in combination with $\ell_2$-minimization sequential inference methods to infer more parameters than just the birth and the death rates in the Lotka-Volterra process introduced in Section \ref{sec:model-LV}. The parameters to be inferred are $\{\delta_S, \delta_R, \gamma_S, \gamma_R, \sigma_S, \sigma_R, r_S, r_R, K_S, K_R, \alpha_S, \alpha_R\}$. We use a sequential inference method to infer each of these parameters. 

\begin{algorithm}[H]
\caption{Summary of Sequential Inference Procedure}\label{alg:sequential-inference}
\begin{enumerate}
    \item Step 1: Infer the total birth and death rates, $\lambda_k^i$ and $\mu_k^i$, from monoculture time series data using the method described in Section \ref{sec:inferencealgorithm_general}. Here, index $k$ denotes cell type $k$, and index $i$ denotes bin $i$.
    \item Step 2: Infer intraspecies parameters 
    $\{\delta_k, r_k, K_k, \gamma_k\}$  via $\ell_2$-minimization, where we minimize the difference between the true and estimated birth and death rates over the parameter set $\{\delta_k, r_k, K_k, \gamma_k\}$.
    \item Step 3: Infer the total birth and death rates, $\lambda_k^{i,j}$ and $\mu_k^{i,j}$, from coculture time series, using the procedure in Section \ref{sec:inferencealgorithm_general}. Here, index $k$ denotes cell type $k$, and index $(i,j)$ denotes bin $(i,j)$.
    \item Step 4: Infer interspecies parameters $\{\sigma_k, \alpha_k\}$ 
    via $\ell_2$-minimization, where we minimize the difference between the true and estimated birth and death rates over the parameter set $\{\sigma_k, \alpha_k\}$. 
\end{enumerate}
\end{algorithm}
\paragraph{Step 1: Inference of monoculture birth and death rates} In the first step, we use the dataset of monoculture data for $S$ and $R$ separately to infer the total birth and death rates at the points {$\{(N_S^i, 0)\}$ and $\{(0,N_R^j)\}$}, respectively. 
In particular, because the dataset includes only monoculture data, the grid blocks are one dimensional and divide up each of the axes. We used a constant grid size $\Delta x$. 
We analyze the error associated with different values of $\Delta x$ in Section \ref{sec:numericalvalidation}. 
The result of this step is a collection of approximated birth and death rates at these population size points, {$\{\lambda_S^i = \lambda_S(N_S^i, 0)\}, \{\mu_S^i = \mu_S(N_S^i, 0)\}, \{\lambda_R^j = \lambda_R(0, N_R^j)\}$ and $\{\mu_R^j = \mu_R(0,N_R^j)\}$}.

\paragraph{Step 2: Inference of monoculture LV-specific parameters for each species} Next, we infer the parameters intrinsic to each subpopulation $\{\delta_S, r_S, K_S, \gamma_S\}$ and $\{\delta_R, r_R, K_R, \gamma_R\}$ by minimizing $\ell_2$ error of these parameters from the birth and death rates found in Step 1. 
Let $\btheta \in [0,\infty) \times [0,\infty) \times [0,\infty) \times [0,1]$ be the parameter vector.
This space is chosen because we minimize over possible $\btheta$ values to get $\btheta^*_S = (\delta_S, r_S, K_S, \gamma_S)$ and $\btheta^*_R = (\delta_R, r_R, K_R, \gamma_R)$. Therefore, we find 
\begin{align}
    \btheta^*_S &= \argmin_{\theta} \sum_i |\lambda_S^i - b^{mono}(N_S^i;\theta)|^2 + |\mu_S^i - d^{mono}(N_S^i; \theta)|^2 \\
    \btheta^*_R &= \argmin_{\theta} \sum_j |\lambda_R^j - b^{mono}(N_R^j;\theta)|^2 + |\mu_R^j - d^{mono}(N_R^j; \theta)|^2, 
\end{align}
where, for {$\btheta = (\delta_\theta, r_\theta, K_\theta, \gamma_\theta)$,} 
\begin{align}
    b^{mono}(m; \btheta) &= (1+\delta_\theta)r_\theta m - \gamma_\theta \frac{r_\theta}{K_\theta} m^2 \label{eq:bmono} \\
    d^{mono}(m; \btheta) &= \delta_\theta r_\theta m + (1-\gamma_\theta)\frac{r_\theta}{K_\theta} m^2. \label{eq:dmono} 
\end{align}
Up to this point, we only use monoculture data. 
We assume that these parameters do not change when the subpopulations evolve together; 
interspecies interactions are reflected only through parameters $\{\sigma_S, \sigma_R, \alpha_S, \alpha_R\}$, which are inferred in Step 4 below. 

\paragraph{Step 3: Inference of coculture birth and death rates }
In this step, we infer the birth and death rates at grid block midpoints using coculture data, again using the technique described in Section \ref{sec:inferencealgorithm_general}. We infer {$\{\lambda_S^{i,j} = \lambda_S(N_S^i, N_R^j)\}, \{\mu_S^{i,j} = \mu_S(N_S^i, N_R^j)\}, \{\lambda_R^{i,j} = \lambda_R(N_S^i, N_R^j)\}$ and $\{\mu_R^{i,j} = \mu_R(N_S^i, N_R^j)\}$} at the points $\{(N_S^i, N_R^j)\}$. We use the same constant $\Delta x$ grid size for steps 1 and 3, but in this step, each grid block is two dimensional. 

\paragraph{Step 4: Inference of LV-specific parameters for interspecies interaction  }
In the final step, we infer parameters $\{\sigma_S, \sigma_R, \alpha_S, \alpha_R\}$ using $\ell_2$ error minimization. We let $\btheta = (\theta_1, \theta_2) \in [0,1] \times \R$. This space is chosen so that minimizing will give $\btheta^*_{S_{int}} = (\sigma_S, \alpha_S)$ and $\btheta^*_{R_{int}} = (\sigma_R, \alpha_R)$. 
\begin{align}
    \btheta^*_{S_{int}} &= \argmin_{\theta} \sum_{i,j} |\lambda_S^{i,j} - b^{co}(N_S^i, N_R^j;\theta^*_S, \theta)|^2 + |\mu_S^{i,j} - d^{co}(N_S^i, N_R^j;\theta^*_R, \theta)|^2 \\
    \btheta^*_{R_{int}} &= \argmin_{\theta} \sum_{i,j} |\lambda_R^{i,j} - b^{co}(N_S^i, N_R^j;\theta^*_R,\theta)|^2 + |\mu_R^{i,j} - d^{co}(N_S^i, N_R^j; \theta^*_R, \theta)|^2, 
\end{align}
where, for $\btheta^* = (\delta^*, r^*, K^*, \gamma^*)$ and $\btheta = (\sigma_\theta, \alpha_\theta)$,
\begin{align}
    b^{co}(m_1, m_2; \theta^*, \theta) &= (1+\delta^*)r^*m_1 - \gamma^* \frac{r^*}{K^*} m_1^2 - \sigma_\theta\alpha_\theta \frac{r^*}{K^*} m_1 m_2 \label{eq:bco}\\
    d^{co}(m_1, m_2; \theta^*, \theta) &= \delta^*r^*m_1 + (1-\gamma^*) \frac{r^*}{K^*} m_1^2 + (1-\sigma_\theta)\alpha_\theta \frac{r^*}{K^*} m_1 m_2. \label{eq:dco}
\end{align}
This gives inferred values for the final four parameters in the model. With this method, all parameters are able to be inferred; see Section \ref{sec:inference-identifiability} for further discussion of identifiability of the parameters. 

\subsection{Numerical Validation}\label{sec:numericalvalidation}

\begin{figure}[!b]
    \centerline{\footnotesize (a) monoculture: \hspace{10.5cm} }
    \vspace{.0cm}
    \centerline{ \footnotesize $\gamma_S   = \gamma_R = 0$, $\sigma_S = \sigma_R = 0$  \hspace{9cm} }         
    \vspace{.0cm}
    \centerline{ 
    \includegraphics[width=15cm]{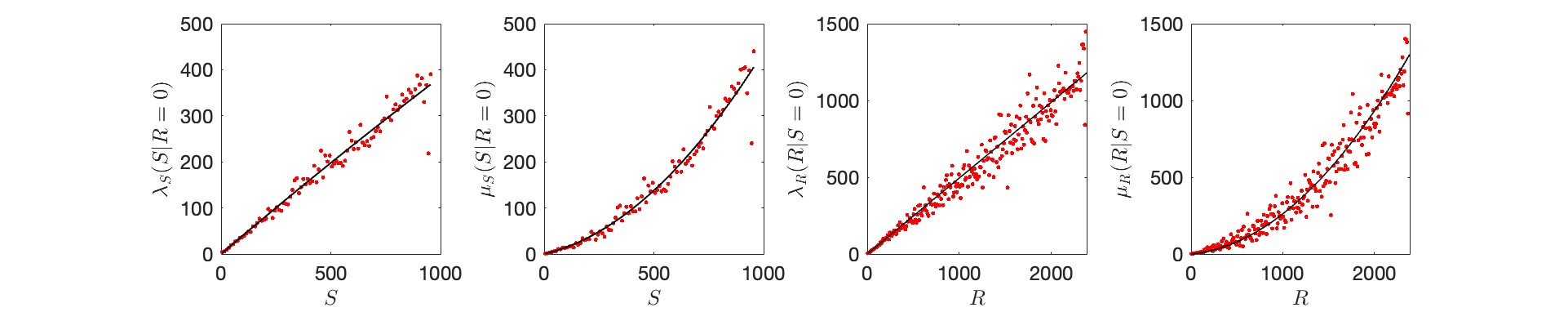}
\hspace{1cm}
}
    \vspace{.0cm}
    \centerline{ \footnotesize $ \gamma_S = \gamma_R = 1$ , $  \sigma_S = \sigma_R = 1$  \hspace{9cm} }
    
    \vspace{.0cm}
    \centerline{ 
\includegraphics[width=15cm]{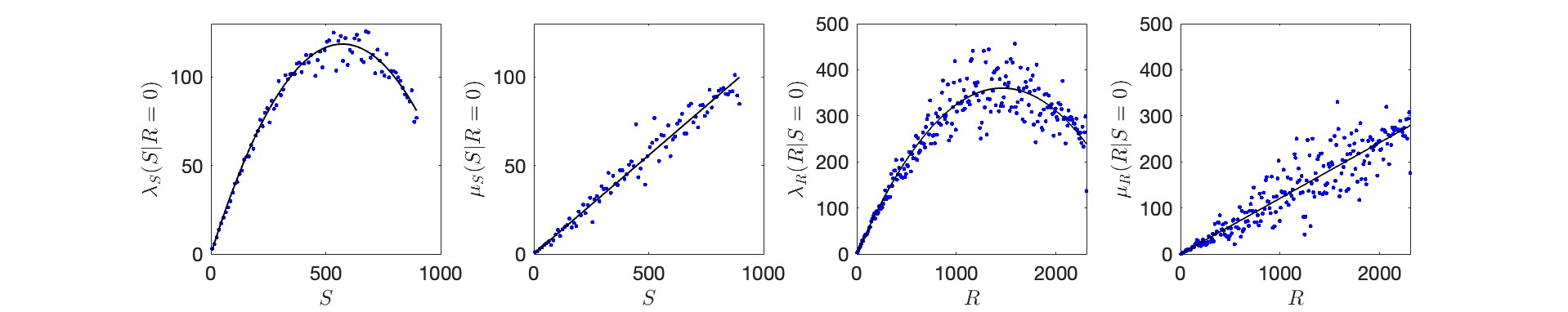}
 \hspace{1cm}
}
    \centerline{\footnotesize (b) coculture: \hspace{11cm} }
    \centerline{ \footnotesize $\gamma_S = \gamma_R = 0$, $\sigma_S = \sigma_R = 0$  \hspace{4.1cm} $\gamma_S = \gamma_R = 1$, $\sigma_S = \sigma_R = 1$ \hspace{0.3cm} }
    \centerline{ 
\includegraphics[width=7cm]{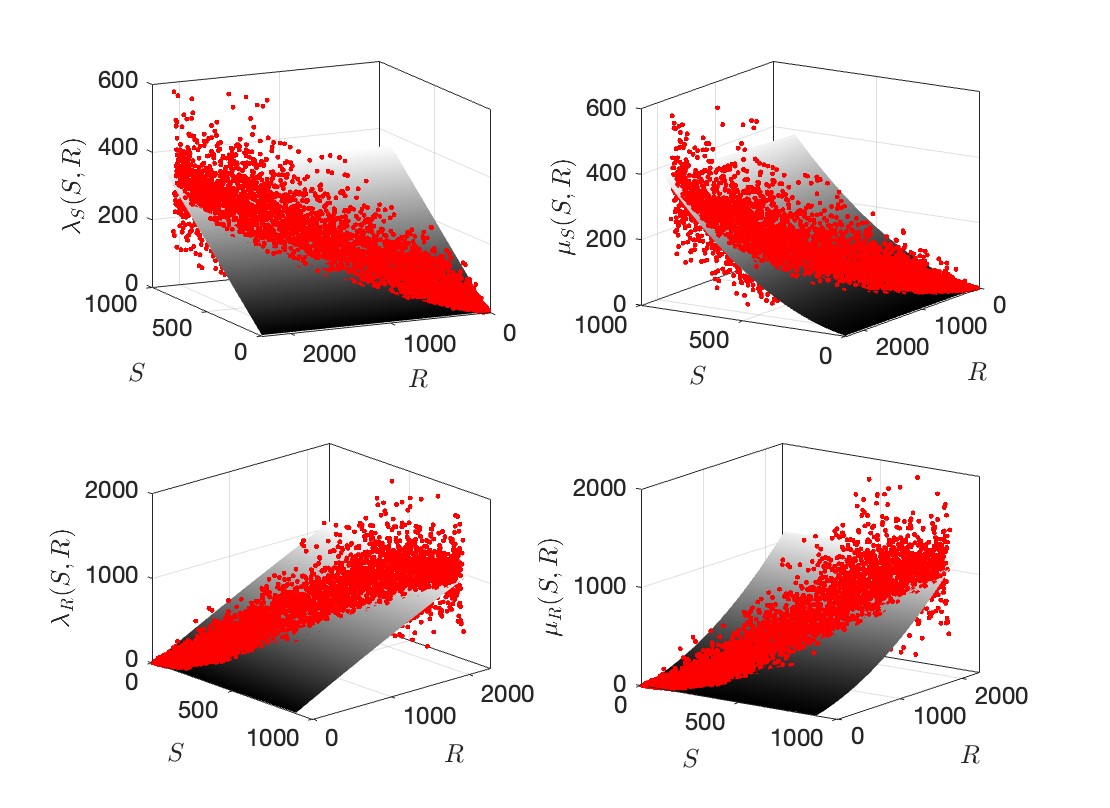}
\includegraphics[width=7cm]{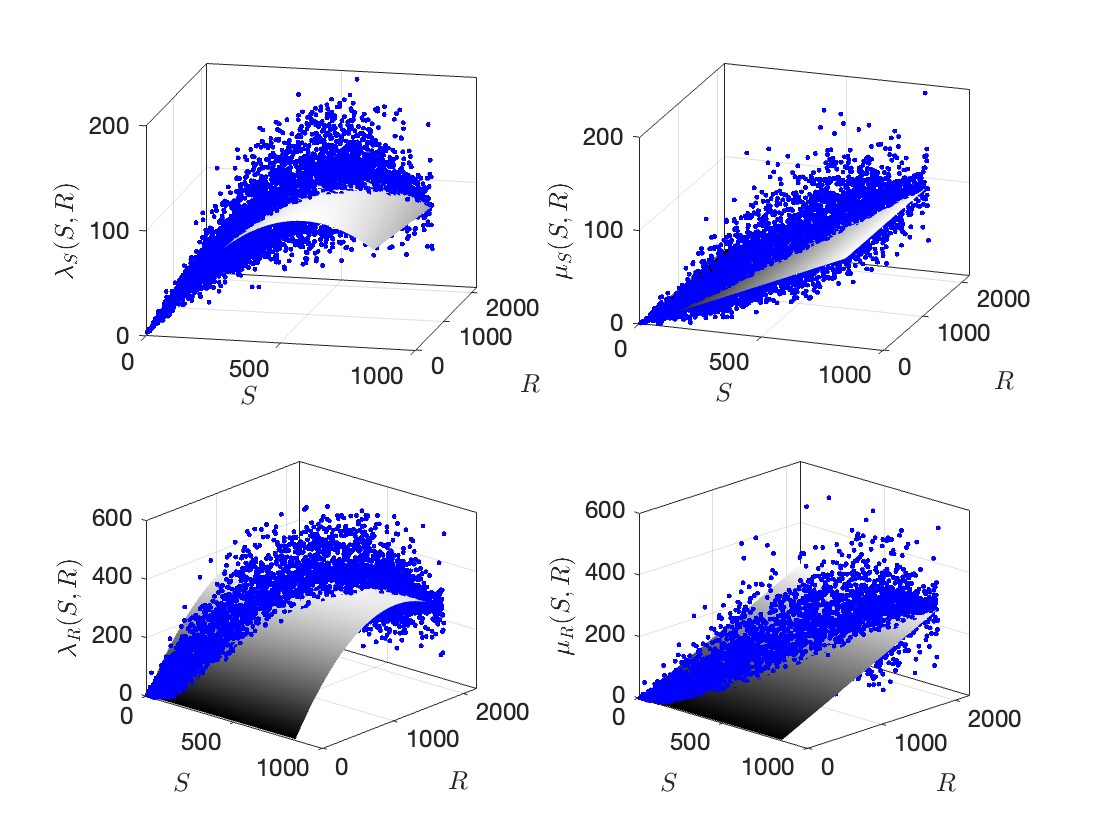}
\hspace{0cm}
}
\caption{Total birth rate $\lambda_k^{i,j}$ and death rate $\mu_k^{i,j}$ (dots) estimated by the method in Section \ref{sec:inferencealgorithm_general}, and the fitted birth and death rate functions, $b^{mono}_k(N_k;\theta)$ and $d^{mono}_k(N_k;\theta)$ for (a) monoculture and $b^{co}_k(N_1,N_2;\theta)$ and $d^{co}_k(N_1,N_2;\theta)$ for (b) coculture, for $k=S,R$, determined by the method in Section \ref{sec:inferencealgorithm_LV}. 
The estimated rate functions $b_k$ and $d_k$ exhibit the correct linear/nonlinear behavior.
When $\gamma_S=\gamma_R=0,\, \sigma_S=\sigma_R=0$ (red), the birth rates are linear functions of $S$ and $R$, while death rates are nonlinear functions of $S$ and $R$. On the other hand, when $\gamma_S=\gamma_R=1,\, \sigma_S=\sigma_R=1$ (blue), the death rates are linear functions, and the birth rates are nonlinear functions.}
    \label{fig:birthdeathrates}
\end{figure}

In this section, we validate the proposed inference method on simulated data, comparing the inferred parameters with the true simulation parameters. In particular, we analyze the inference errors in terms of number of time series $M$, discretized time step $\Delta t$, and grid size $\Delta x$. 
Unless mentioned otherwise, the default values of the inference method parameters are chosen as follows: $M=100$,   $\Delta x = 10$,   $\Delta t = 0.1$. The bin exclusion threshold is set at $100$, where bins containing less than 100 data points are excluded. Since the subpopulation carrying capacities in our simulated data are on the order of $10^3$, the choice of grid size $\Delta x = 10$ divides the time series into $O(10^2)$ monoculture bins and $O(10^4)$ coculture bins. The final time of the stochastic simulation is $T=100$; thus, $\Delta t = 0.1$ yields 1000 discretized time points. For the other model parameters, we focus on the specific values associated with the cancer cell line PC3. Parameters of the PC3 cell line can be found in Table \ref{table:paramfit}.

In Figure \ref{fig:birthdeathrates}, we show the estimated birth and death rates as functions of subpopulation sizes from the monoculture and coculture time series of the PC3 cell line. We simulate the two extreme cases: where intraspecies and interspecies interactions are regulated solely by the death process ($\gamma_S = \gamma_R = 0$, $\sigma_S = \sigma_R = 0$) (red) and where they are regulated solely by the birth process ($\gamma_S = \gamma_R = 1$, $\sigma_S = \sigma_R = 1$) (blue). 
In the case where $\gamma_S = \gamma_R = 0$ and $\sigma_S = \sigma_R = 0$, the death rates are nonlinear functions of subpopulation sizes $S$ and $R$, while the birth rate functions are linear in $S$ and $R$. In the monoculture data, this results in a line, and in the coculture data, we see a plane. 
On the other hand, when the regulation only affects the birth process ($\gamma_S = \gamma_R = 1$ and $\sigma_S = \sigma_R = 1$), the birth rate functions are nonlinear, and the death rate functions are linear. 

The red and blue dots in Figure \ref{fig:birthdeathrates} are the rates estimated from the time series statistics, mean and variance, in each bin. 
Note that not all bins have the rates estimated, especially in the coculture data; the rates are not estimated in bins without enough data points. 
Using the red and blue dots, the lines in Figure \ref{fig:birthdeathrates}(a) are the estimated monoculture birth/death rates, and the gray surfaces in Figure \ref{fig:birthdeathrates}(b) are the coculture birth/death rates. 
We note that the fitted lines and surfaces of the birth and death rate functions are consistent with the exact rate functions.

\begin{figure}[!htb]  
\centerline{
\includegraphics[width=12cm]{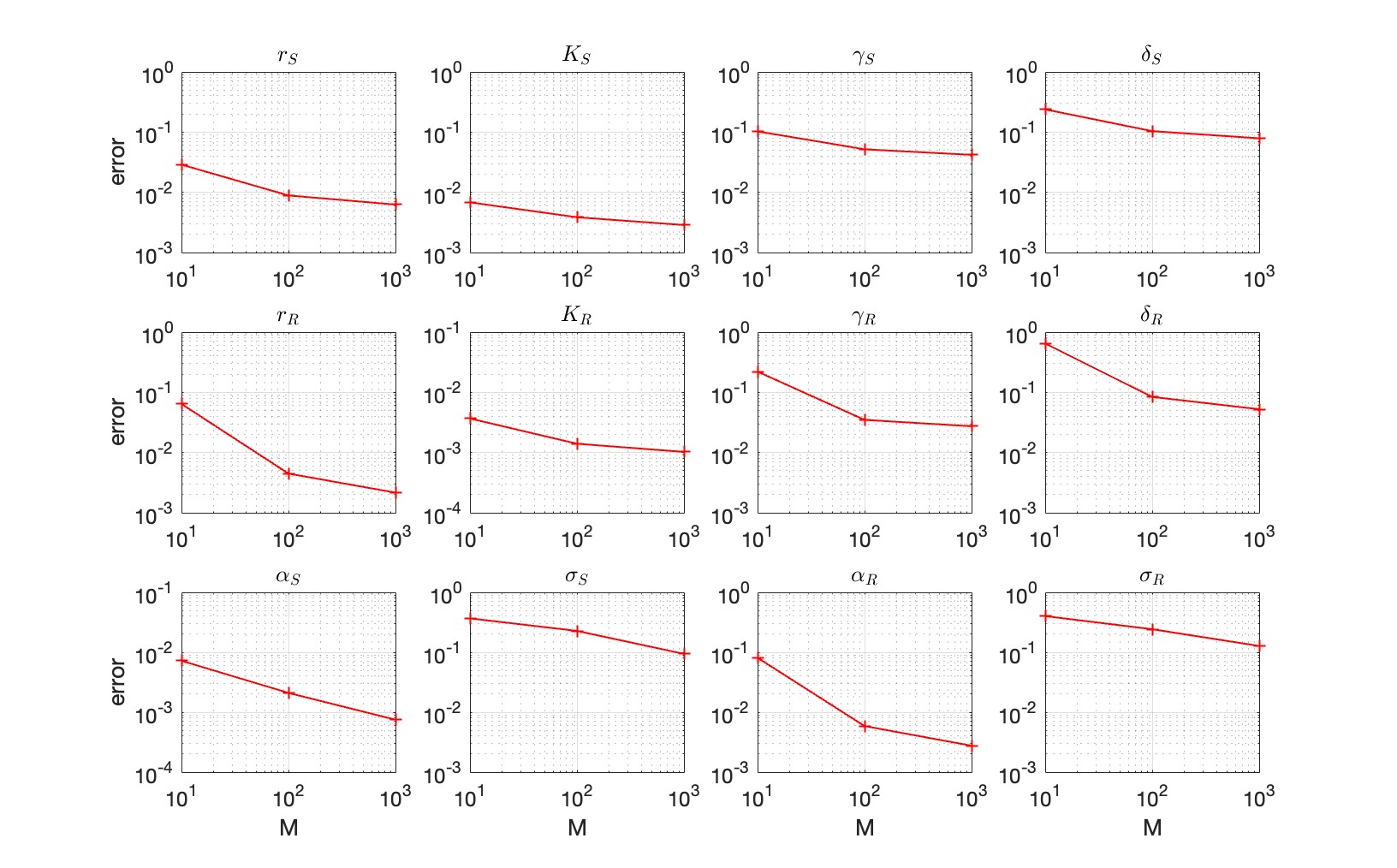}
}
\caption{
Relative error of PC3 parameter estimation with respect to number of time series $M = 10, 100, 1000$. The error decreases as the number of time series increases.
}
    \label{fig:error_S}
\end{figure}

\begin{figure}[!htb]  
\centerline{
\includegraphics[width=12cm]{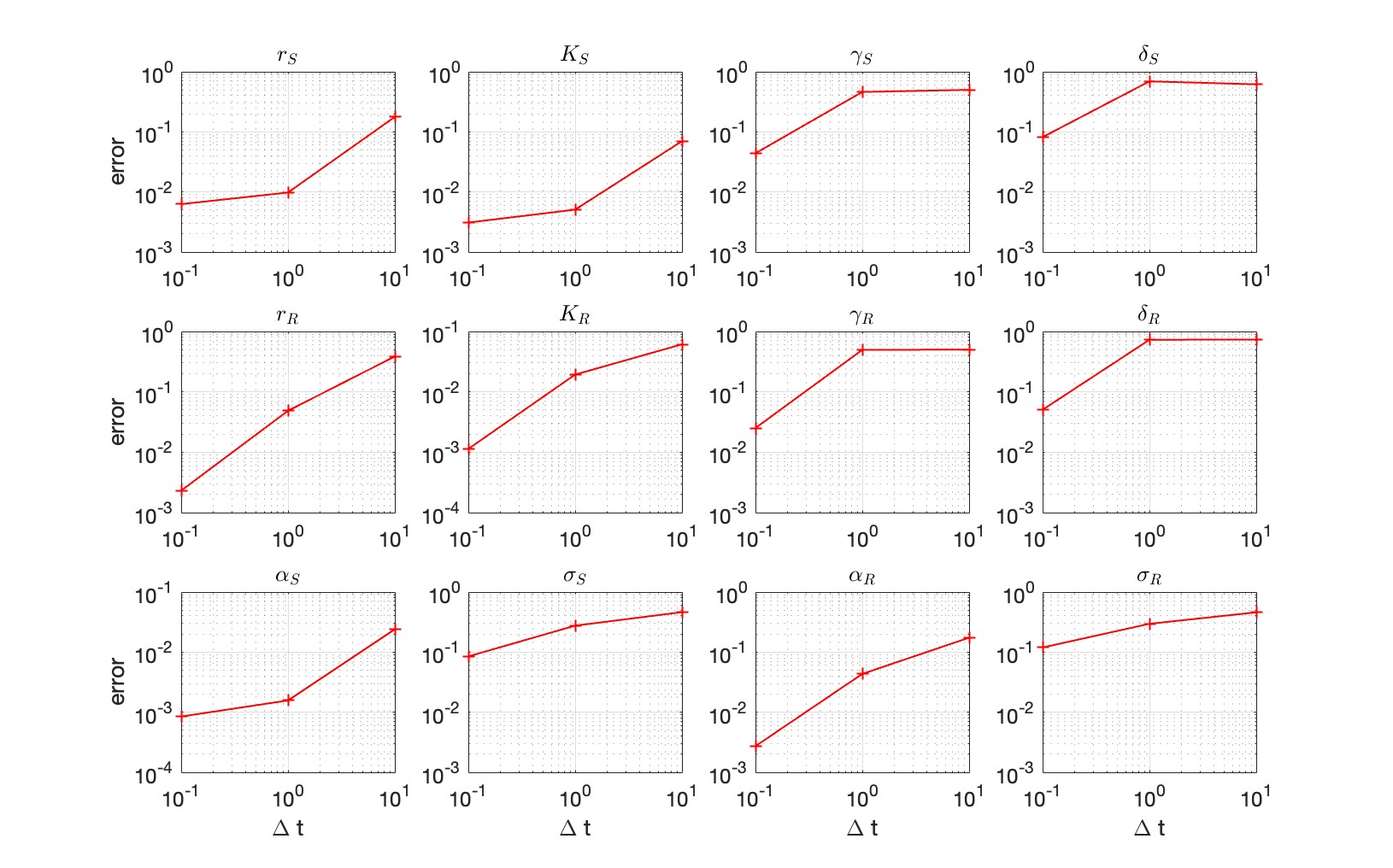}
}
\caption{Relative error of PC3 parameter estimation with respect to time step size $\Delta t = 0.1, 1, 10$. The error decays as $\Delta t$ decreases.
}
    \label{fig:error_dt}
\end{figure}

\begin{figure}[!htb]  
\centerline{
\includegraphics[width=12cm]{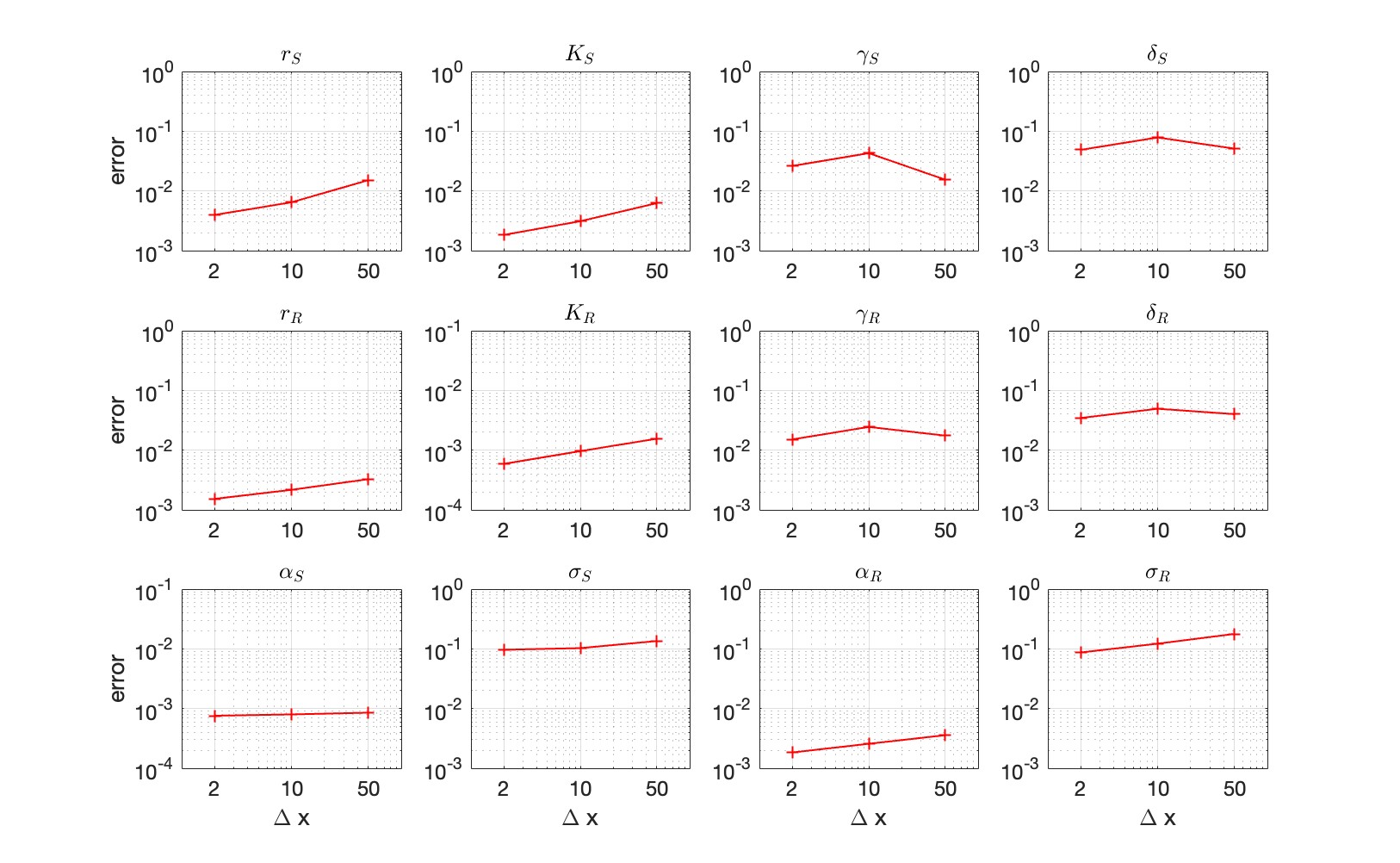}
}
\caption{Relative error of PC3 parameter estimation with respect to grid size $\Delta x = 2,10,50$. The error is less sensitive to the grid size than other inference parameters.
}
    \label{fig:error_dS}
\end{figure}

%
In Figures \ref{fig:error_S}--\ref{fig:error_dS}, we study the error in the estimated parameter values with respect to the hyper-parameters of the inference methods. In particular, we study the convergence in terms of the number of time series $M$, time step $\Delta t$, and grid size $\Delta x$. The parameter set of the PC3 cell line is considered, and the relative $\ell_2$ error among the 100 inference trials is computed as $\frac{1}{100} \sqrt{ \sum_{i=1}^{100} ((\theta^{true} - \tilde{\theta}^{(i)}) / \theta^{true} )^2 }$, where $\tilde{\theta}^{(i)}$ is the estimated parameter in the $i$-th inference trial. 

Figure \ref{fig:error_S} shows the relative error of the parameters while increasing the number of time series $M$ for $M = 10, 100$, and $1000$. The error monotonically decreases as $M$ increases from 10 to 1000 and confirms that having more data improves the accuracy of the inferred parameters. 
Figure \ref{fig:error_dt} shows the error with respect to $\Delta t$, the time step of stochastic simulation. The shown results are for $\Delta t = 0.1, 1$, and $10$, and the accuracy improves as $\Delta t$ decreases. Due to stochasticity in the time series, using a time step that is too small makes the rate inference less accurate. Moreover, it shows that using a large time step such as $\Delta t = 10$ significantly deteriorates the accuracy. 
In Figure \ref{fig:error_dS}, the grid sizes $\Delta x= 2, 10$, and $50$ are tested. The error seems to be the least sensitive to $\Delta x$ as compared to the other tested hyper-parameters $M$ and $\Delta t$. However, as mentioned above, the choice of grid size will influence which grid blocks have enough population points to compute the mean and variance; here, this cutoff is set at 100. 


\subsection{Parameter Identifiability through Stochasticity}\label{sec:inference-identifiability}
In the proposed inference method, using the stochasticity via the statistics of the time series data is critical for inferring the parameters $\gamma_k$, $\sigma_k$, and $\delta_k$ related to the birth and death processes. If only the deterministic population growth curves are used in model calibration, the data can be fitted to the Lotka-Volterra model in Equations \eqref{eqn:dSdt}--\eqref{eqn:dRdt}. 
The structural identifiability of the Lotka-Volterra model has been established in \cite{Remien, Greene, cho2023designing}. 
Thus, the Lotka-Volterra model parameters $r_k$, $K_k$, and $\alpha_k$ are structurally and practically identifiable with an appropriate amount of data. 
However, the weight parameters that distinguish the effect of interaction on birth and death cannot be identified using deterministic time series data since the Lotka-Volterra model does not distinguish whether the net growth rate of the population is contributed by the birth or death process. 
Thus, the statistics of the time series, in particular the mean and variance of the time series, are essential to estimate the parameters $\gamma_k$, $\sigma_k$, and $\delta_k$. 

\begin{figure}[H]
   \centerline{ 
\includegraphics[width=14cm]{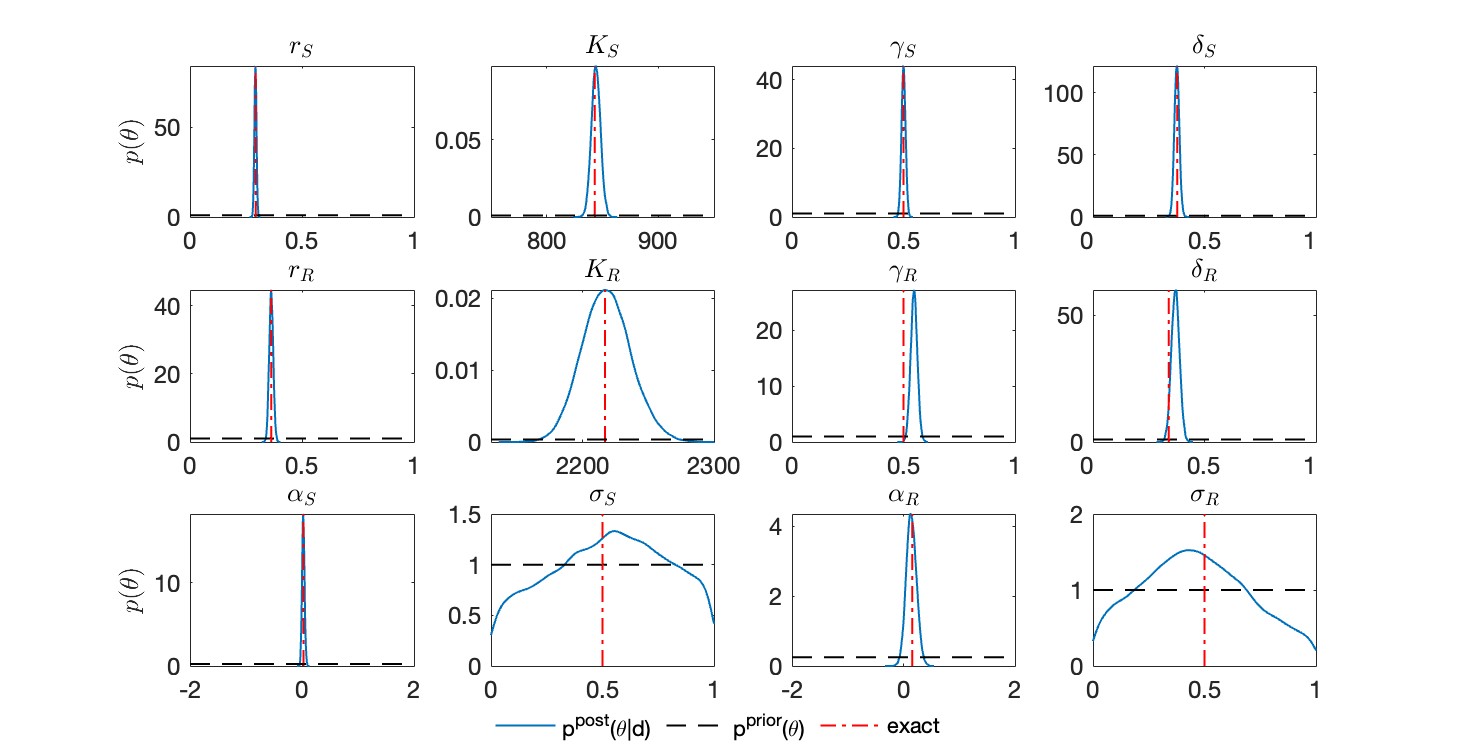}
}
\caption{ Prior (dash) and posterior (line) probability distributions of the parameters, computed using a Markov chain Monte Carlo method (MCMC). The posterior distribution of each parameter is a unimodal function centered around the true parameter value (vertical dash-dot line). This suggests that our inference method is practically identifiable, and it is able to uniquely estimate the parameters that distinguish the effects of the birth and death processes. }
    \label{fig:mcmc}
\end{figure}

We further study the practical identifiability of parameters by our inference algorithm using Bayesian calibration. The goal of practical identifiability is to determine whether model parameters can be inferred from potentially noisy data via model calibration. Practical identifiability addresses difficulty in inferring parameter values due to measurement errors, model discrepancy, or an experimental design that is unable to naturally activate certain parameters. These issues depend on the quality, design, and availability of the data, rather than the model structure. In this section, we say that a parameter is practically identifiable if the posterior distribution of a parameter $\theta$ given a dataset $D = \{ \lambda_k^{i,j}, \mu_k^{i,j} \}$, 
\begin{align}
p^{post}(\theta|D) = \frac{p(D|\theta) p^{prior}(\theta)}{p(D)} \propto p(D|\theta) p^{prior}(\theta)
\end{align}
is unimodal, exhibiting a clear and unique optimum. Otherwise, the parameter is interpreted as practically non-identifiable, in the sense that multiple values of the input may yield the same model output. We compute the posterior distribution using the Metropolis Hastings algorithm, a Markov chain Monte Carlo (MCMC) method using the delayed rejection adaptive metropolis (DRAM) algorithm developed in \cite{Haario}. The prior distribution $p^{prior}(\theta)$ of each parameter is taken to be a uniform distribution on the range in Table \ref{table:modelparamlist}, assuming that we have no prior knowledge about the parameter other than its range. We choose a Gaussian distribution for the likelihood function $p(D|\theta)$, 
\begin{align}
p(D|\theta) = \prod_{i\in I} \frac{1}{\sqrt{2 \pi \sigma^2 }} {\exp}\left( \frac{ (D_i - F(N_i;\theta) )^2 }{2 \sigma^2 } \right)
\end{align}
where $F(N_i;\theta)$ are the birth and death rate functions in Equations (\ref{eq:bmono}--\ref{eq:dmono}, \ref{eq:bco}--\ref{eq:dco}) to be fitted, which depend on parameters $\theta$. This form of the likelihood assumes an additive error that is Gaussian.  
We run 8 independent runs of MCMC with a burn-in period of 10000 steps, followed by 50000 iterations. The initial points are chosen using Latin hypercube samples \cite{helton2003latin}. 
The convergence of the MCMC algorithm is checked via $\hat{R}$ \cite{gelman1995bayesian} and the value is found to be 1.003 {(significantly below the typical tolerance 1.05)}. 
\begin{table}[!htbp]
\begin{tabular}[c]{c|c|cc}
\toprule 
         &  \multicolumn{3}{c}{PC3 Cell line}   
         \\ 
\midrule
 Parameter & true  & Median & 90\% CI \\ 
\midrule 
 $r_S$ & 0.293 &   0.2922 & $[0.2844,   0.3000 ]$ 
\\ 
 $K_S$ & 843 & 844.1  &    $[ 837.5,    850.9  ]$   
\\ 
 $\delta_S$ & 0.3784 & 0.3763  &   $[ 0.3581,   0.3944 ]$  
\\ 
 $\gamma_S$ & 0.5 & 0.5004  &   $[ 0.4854,    0.5155  ]$  
\\ 
 $r_R$ & 0.363 &  0.3627  &   $[ 0.3477,    0.3778  ]$  
\\ 
 $K_R$ & 2217 &  2218  &    $[ 2187,   2249  ]$  
\\ 
 $\delta_R$ & 0.3396 &  0.3701 & $[ 0.3403,   0.4003 ]$   
\\ 
 $\gamma_R$ & 0.5 & 0.5474  &   $[  0.5228,    0.5722  ]$  
\\ 
 $\alpha_S$ & 0.027 & 0.02269   &  $[ -0.01431,    0.06126 ]$  
\\ 
 $\sigma_S$ & 0.5 & 0.5191  &   $[ 0.07912,    0.9360  ]$  
\\ 
 $\alpha_R$ & 0.159 & 0.1384  &   $[-0.0068,    0.2896  ]$      
\\ 
 $\sigma_R$ & 0.5 &  0.4578 &   $[ 0.07214,    0.8870  ]$   
\\ 
 \botrule 
\end{tabular}
\caption{Estimated parameters compared to the true parameter values. Considering the median and 90\% confidence interval drawn from the posterior distribution, the inferred parameters are close to the true value in the PC3 cell line \cite{paczkowski2021}. See Appendix \ref{secA1} for additional information.}
\label{table:paramfit}
\end{table}
In Figure \ref{fig:mcmc}, the prior distribution $p^{prior}(\theta)$ and posterior distributions $p^{post}(\theta|D)$ are plotted. The posterior distributions of all parameters are unimodal around the true parameter values, indicating accuracy of our inference method. 
In addition, Table \ref{table:paramfit} shows the statistics of the fitted parameter values for the PC3 cell line. The median and 90\% confidence interval drawn from the posterior distribution of each parameter are close to the true parameter value. This suggests that the proposed inference method is practically identifiable, and it can estimate the birth and death process parameters $\gamma_k$, $\sigma_k$, and $\delta_k$. 
We note that our inference method and results include a mix of frequentist and Bayesian methods. While the statistical properties of the method require further theoretical validation, simulation results demonstrate its practical effectiveness. 

\section{Conclusions and Discussion}\label{sec:conclusions}
In this paper, we study the significance of ecological interactions and separation of birth and death dynamics in stochastic heterogeneous populations via general birth-death processes.
In the Lotka-Volterra model, we see several notable impacts on the way the subpopulations evolve when changing how the intraspecies and interspecies interactions affect the birth and death rates. Moving the intraspecies competition to the birth rate decreases the range of the population sizes seen after a long enough period of time. Changing the interspecies interaction from the birth rate to the death rate does not have as big of an impact on the final subpopulation sizes. This is true for both the PC3 and the DU145 cell line parameter values. 
The placement of the interspecies interaction in the birth versus the death rate has a much bigger impact on the survival probability of the resistant population. The type of interaction determines whether the survival probability increases or decreases as the interspecies interaction moves from the death rate to the birth rate. The intraspecies competition has less of an impact on the survival probability. This is likely because when the resistant subpopulation size is small and therefore has a reasonable chance of extinction, the size of the intraspecies term is dwarfed by the size of the interspecies interaction term. Once the resistant population has grown to a size where the intraspecies term is comparable to the interspecies term, the probability of the population going extinct is small enough to be negligible on the time scale of the calculations. 

We introduced an inference method for disambiguating the birth and the death rate in a general birth-death process with $n$ distinguishable subtypes. Our inference method requires several different time series of the counts of each subpopulation over time. This method introduced in Section \ref{sec:inferencealgorithm_general} can be used even if the functional forms of the birth rate and the death rate are unknown. The main idea behind the method is to utilize the properties of the stochastic system to infer the rates. Here, we use the mean and variance, which is similar to the idea of the method of moments inference technique, which uses moments of the data to infer different parameter values. One could consider using a maximum likelihood approach to infer these parameters as well. However, because the dataset does not contain the birth and death counts, only the population counts at different time points, the distribution needed for a maximum likelihood analysis is that of the difference between two Poisson random variables, known as the Skellam distribution. An analysis comparing the method of moments to maximum likelihood techniques in this setting was done by Alzaid et al.~2010 \cite{alzaid}. 

We then demonstrated this inference method in the context of the Lotka-Volterra birth-death process. In this example, we pair the inference method for disambiguating birth and death with a sequential $\ell_2$-minimization inference technique to infer the parameters in the birth and death rate functions. The order of the error was discussed as a function of the number of time series, the size of the time step $\Delta t$, and the grid size $\Delta x$. In this case, we are able to identify all parameters through the combination of these methods. The identifiability of the birth and death rates was a result of the underlying stochasticity in the process.

There are several open questions and extensions which remain related to expansion of the theory and advancing the applications. In terms of the theory, the accuracy of the method presented here relies on data collected with a small $\Delta t$ time step between data points. This is realistic for in-silico data and even for some in-vitro data sources \cite{roshan2015dynamic} but is unrealistic for in-vivo studies. Expanding the method to be accurate for large $\Delta t$ time steps will require additional insights. Additionally, the method requires that the subtypes be able to be counted separately, but it is of interest to consider a way to relax this condition so that inference is possible in systems where the subtypes are indistinguishable. This would allow for broader application of the method. A limitation of our work is that we do not consider spatial effects in the population interactions, and future work includes incorporating the birth-death inference step into a spatial model. The applications of the technique would be more involved because of the additional variables to consider, such as spatial interaction and spatially correlated noise in the data. Considering phenotype switching in combination with interactions is another future direction. We also propose to consider different inference strategies for the Lotka-Volterra model parameters, such as inferring all model parameters at once \cite{cho2023designing} instead of sequentially, with an aim to understand the difference in the inference error in these situations. 

One of the first applications of this method would be to consider the addition of drug effects to the model. One could consider the question of which dosing regimes are optimal based on how the subpopulation interactions are occurring (birth versus death) and how the drug is impacting the subpopulations (birth versus death). Applying the method to in-vitro data sets is another step which would further validate the method and expand the systems to which the inference method can be applied.

\backmatter

\section*{Declarations}

\begin{itemize}
\item Conflict of interest/Competing interests: The authors declare that they have no known competing financial interests or personal relationships that could have appeared to influence the work reported in this paper.

\item Data availability: There is no experimental or clinical data. Simulated data can be reproduced using the code below.
\item Code availability: Code can be found at \url{https://github.com/lhuynhm/inference_heterogeneous_interactions_birthdeath} 
\end{itemize}


\bigskip





\begin{appendices}




\newpage
\section{Reference for Parameters}\label{secA1}
\begin{table}[!htbp]
\begin{tabular}[c]{c|c|c|c|c}
\toprule 
 Parameter & PC3 Cell line  & DU145 Cell line & Reference & Range \\
\midrule 
 $r_S$ &  0.293 &   0.306  &  \cite{paczkowski2021}\hphantom{$^*$} & [0, 1]  \\
 $K_S$ &  843    & 724  & \cite{paczkowski2021}$^*$ & [1, $10^4$] \\
 $\delta_S$ &  0.3784     & 0.3784 & \cite{paczkowski2021}$^*$  & [0, 1] \\
 $r_R$ &  0.363   & 0.21 & \cite{paczkowski2021}\hphantom{$^*$}   & [0, 1]  \\
 $K_R$ &  2217  & 1388 & \cite{paczkowski2021}$^*$  & [1, $10^4$]  \\
 $\delta_R$ &  0.3396   & 0.3396  & \cite{paczkowski2021}$^*$  & [0, 1] \\
$\alpha_S$ &  0.027  & ${-0.501}$ & \cite{paczkowski2021}\hphantom{$^*$}  & [$-$2, 2] \\
 $\alpha_R$ &  0.159   & 0.221 & \cite{paczkowski2021}\hphantom{$^*$} & [$-$2, 2]\\ 
 $\gamma_S$ & [0,1] & [0,1]  & $\star$  & [0, 1]\\ 
 $\gamma_R$  &  [0,1] & [0,1]  & $\star$  & [0, 1]\\ 
 $\sigma_S$ &    [0,1] & [0,1]  & $\star$   & [0, 1] \\ 
 $\sigma_R$ &   [0,1] & [0,1] & $\star$  & [0, 1] \\ 
 \botrule 
\end{tabular}
\caption{Model parameters of prostate cancer cell lines, PC3 and DU145, their values and references. Most of the Lotka-Volterra model parameters are directly from \cite{paczkowski2021}. The parameters with \cite{paczkowski2021}$^*$ are the values recalculated from the cellular automata model developed in \cite{paczkowski2021} to match the experimental data. The birth and death weight parameters with $\star$ are the values studied in this paper. Parameter ranges used in the inference method are shown.}
\label{table:modelparamlist}
\end{table}

\end{appendices}


\input{output.bbl}

\end{document}

%% file: macros.tex
\usepackage{graphicx}%
\usepackage{multirow}%
\usepackage{amsmath,amssymb,amsfonts}%
\usepackage{amsthm}%
\usepackage{mathrsfs}%
\usepackage[title]{appendix}%
\usepackage{xcolor}%
\usepackage{textcomp}%
\usepackage{manyfoot}%
\usepackage{booktabs}%
\usepackage{algorithm}%
\usepackage{algorithmicx}%
\usepackage{algpseudocode}%
\usepackage{listings}%
\usepackage{float}
\usepackage{ulem}
\usepackage{tikz}
\usepackage{pgfplots}

\usepackage{geometry, hyperref}   
\usepackage{amsmath, tabularx, color, mathtools, array, etoolbox, epsfig, amsthm}
\usepackage{caption}
\usepackage{bbm}
\usepackage{soulutf8}
\usepackage{float}
\usepackage{comment}
\usepackage{graphicx} 
\usepackage{url}
\usepackage{amssymb}
\usepackage{multirow}
\usepackage{subfigure}
\usepackage{bm}
\usepackage{bbm}
\usepackage{placeins}

\newcommand{\mcN}{\mathcal N}

\newcommand{\be}{{\boldsymbol e}}

\newcommand{\bN}{{\boldsymbol N}}
\newcommand{\bt}{{\boldsymbol t}}
\newcommand{\btheta}{{\boldsymbol \theta}}

\newcommand{\R}{\mathbb{R}}

\newcommand{\E}{\mathbb E}

\theoremstyle{definition}

\definecolor{darkgreen}{rgb}{0,0.5,0}
\definecolor{darkred}{rgb}{0.9,0.1,0.1}
\definecolor{purple}{rgb}{0.29,0.0,0.51}

\newcommand{\eb}[1]{{\color{darkgreen}{#1}}}
\definecolor{purple_hc}{rgb}{0.6, 0.2, 0.8}

\newcommand{\linh}[1]{{\color{cyan} Linh: #1}}

\newcommand{\Z}{\mathbb{Z}}

\DeclareMathOperator{\Var}{Var}

\DeclareMathOperator*{\argmin}{arg\,min}
\DeclareMathOperator{\Poi}{Poi}